\RequirePackage[T1]{fontenc}
\documentclass[12pt]{article}

\pdfoutput=1
\usepackage[height=8.85in,width=6.45in]{geometry}
\renewcommand{\baselinestretch}{1.3}

\usepackage[utf8]{inputenc}
\usepackage[svgnames]{xcolor}
\usepackage{graphicx}
\usepackage{appendix}

\usepackage{ascmac}
\usepackage{amsmath}
\usepackage{amssymb}
\usepackage{mathtools}
\usepackage{bm}
\numberwithin{equation}{section}

\usepackage[
	bookmarks=false,
	colorlinks=true,
	citecolor=refcolor,
	linkcolor=refcolor,
]{hyperref}

\usepackage[hang]{caption}
\usepackage[caption=false]{subfig}

\definecolor{refcolor}{rgb}{0.3,0.3,1}
\renewcommand{\thesection}{\arabic{section}}
\renewcommand{\thesubsection}{\thesection.\arabic{subsection}}
\usepackage{braket}
\usepackage{array}
\usepackage{multirow}
\usepackage[hang,bottom]{footmisc}
\usepackage{tikz}
\usetikzlibrary{quantikz}
\usetikzlibrary{snakes}

\usepackage{titlesec}

\titleformat*{\section}{\large\bfseries}
\titleformat*{\subsection}{\bfseries}

\begin{document}

\begin{titlepage}
\begin{center}

\ 

\vspace{20pt}

{\large \bfseries
	Modal analysis 
	on quantum computers via qubitization
}
\renewcommand*{\thefootnote}{\fnsymbol{footnote}}

\vspace{1cm}
Yasunori Lee\footnote{
	\url{lee.y@qunasys.com}
}
and
Keita Kanno\footnote{
	\url{kanno@qunasys.com}
}

\vspace{2pt}

\textit{
	QunaSys,
	Bunkyo, Tokyo, Japan
}

\end{center}

\vspace{10pt}

Natural frequencies and normal modes are basic properties of a structure
which play important roles in analyses of its vibrational characteristics.
As their computation reduces to solving eigenvalue problems,
it is a natural arena for application of quantum phase estimation algorithms,
in particular for large systems.
In this note, we take up some simple examples of (classical) coupled oscillators and show how the algorithm works
by using qubitization methods based on a sparse structure of the matrix.
We explicitly construct block-encoding oracles along the way,
propose a way to prepare initial states,
and briefly touch on a more generic oracle construction for systems with repetitive structure.
As a demonstration, we also give rough estimates of
the necessary number of physical qubits and actual runtime it takes
when carried out on a fault-tolerant quantum computer.

\end{titlepage}

\newpage

\renewcommand{\baselinestretch}{0.6}
\tableofcontents
\thispagestyle{empty}

\newpage

\renewcommand{\baselinestretch}{1.3}
\setcounter{footnote}{0}
\setcounter{page}{1}

\section{Introduction and Summary}

The ultimate target of quantum computation is to solve problems
which are practically intractable by classical computation
due to their size and/or computational time.
Among these problems is computing
eigenvalues 
of a large matrix,
where there is a quantum algorithm called the \emph{quantum phase estimation} (QPE) \cite{Kitaev:1995, CleveEkertMacchiavelloMosca:1998}
which is exponentially efficient compared to any known classical algorithms
(e.g. conjugate gradient method)
in terms of both space and time (gate) complexity.\footnote{
	Strictly speaking,
	one needs to prepare a suitable initial quantum state
	to compute a \emph{specific} eigenvalue,
	which in general cannot be done efficiently and ruins the exponential advantage.
	Here we assume that a desirable initial state can be easily prepared,
	which is the case for the problems in this note ($\to$ Sec.\,\ref{sec:state_preparation}).
}
This task is of great importance as it covers a vast range of applications,
including in particular quantum chemistry/physics problems of
computing the energy of large systems given their Hamiltonians.

To figure out whether the desired QPE can \emph{actually} be carried out within reasonable time,
one must go beyond asymptotic scaling and estimate concrete numbers of qubits and quantum gates (especially non-Clifford ones)
needed to implement the algorithm in a fault-tolerant manner.
To the authors' knowledge, the program along this line was initiated in a full-fledged form by \cite{ReiherWiebeSvoreWeckerTroyer:2016},
employing \emph{Lie-Trotter-Suzuki decomposition}
and resulting in costs small enough to raise hopes but motivating further reductions.
In the past few years, 
a novel powerful method called \emph{qubitization} \cite{LowChuang:2016} has been developed, and
utilizing it significantly reduced the number of necessary non-Clifford gates (at the expense of a moderate number of qubits),
see e.g. \cite{BabbushGidneyBerryWiebeMcCleanPalerFowlerNeven:2018,
	BerryGidneyMottaMcCleanBabbush:2019,
	vonBurgLowHanerSteigerReiherRoettelerTroyer:2020,
	LeeBerryGidneyHugginsMcCleanWiebeBabbush:2020,
	KimLiuPallisterPolRobertsLee:2021,
	SuBerryWiebeRubinBabbush:2021,
	GoingsWhiteLeeTautermannDegrooteGidneyShiozakiBabbushRubin:2022,
	DelgadoCesaresdosReisZiniCamposCurzHernandezVoigtLoweJahangiriMartinDelgadoMuellerArrazola:2022,
	BluntCampsCrawfordIzsakLeonticaMiraniMoylettScivierSunderhaufSchopfTaylorHolzmann:2022,
	IvanovSunderhaufHolzmannEllabyKerberJonesCamps:2022,
	YoshiokaOkuboSuzukiKoizumiMizukami:2022,
	BeverlandMuraliTroyerSvoreHoeflerKliuchnikovLowSoekenSundaramVaschillo:2022,
	RubinBerryMaloneWhiteKhattarDePrinceSicoloKuhnKaicherLeeBabbush:2023,
	ZiniDelgadodosReisCesaresMuellerVoigtArrazola:2023}.
At the time of writing, however,
almost all the literature on application of qubitization-based QPEs has focused on quantum chemistry/physics problems,
relying on decomposition of a target matrix (i.e.~Hamiltonian) into a linear combination of unitaries (LCU).

In this note,
we take up yet another important class of eigenvalue problems, namely \emph{modal analysis},
and see how quantum phase estimation algorithms can be applied for toy problems of (classical) coupled oscillators.
Modal analysis is important since the knowledge of natural frequencies and normal modes of a structure
is indispensable to avoid unwanted resonances leading to collapses or failures,
which makes it appear in various engineering scenes~\cite{AnsysCourse}.
Notably, in the simplest cases, analyses reduce to eigenvalue problems of a sparse matrix,
and thus one can draw on qubitization relying not on a decomposition into LCU but on a sparse structure of the matrix,
which gives them a somewhat different flavor compared to the existing analyses.\footnote{
	This should not be confused with so-called \emph{sparse qubitization}
	which is actually LCU-based qubitization,
	where the (part of) original Hamiltonian is made ``sparse'' by truncating small coefficients.
}
Although there are some previous studies \cite{CampsLinBeeumenYang:2022, SunderhaufCampbellCamps:2023} which seem to be in a similar vein at first sight,
their approach is top-down and motivated by generic well-structured matrices themselves,
while our work is bottom-up and more problem-oriented, which makes the direction orthogonal to some extent.
The toy problems themselves are quite trivial and would provide hardly any practical benefit,
but the resource estimation at least confirms the exponential advantage of quantum computation over its classical counterpart,
and the authors hope that this work would serve as a step toward analyses of realistic uses in more generality:
For example, the oracle construction introduced in this note can be applied to 
efficient simulations of exponentially many coupled oscillators described by \cite{BabbushBerryKothariSommaWiebe:2023}.

\bigskip

The rest of the note is organized as follows.
In Section~\ref{sec:qubitization_review}, we give a brief overview of how to block-encode a generic sparse matrix into a unitary matrix and how to \emph{qubitize}~it.
In Section~\ref{sec:main_analysis}, we consider concrete models of (classical) coupled oscillators and explicitly construct the unitary in terms of quantum circuits, starting from those of underlying oracles.
In Section~\ref{sec:state_preparation}, we propose a way to prepare the input initial state of a QPE algorithm.
Finally in Section~\ref{sec:resource_estimation}, we roughly estimate total resources required to carry out the desired QPE.
Appendix~\ref{appendix:arithmetic} provides elementary-gate-level implementations of various circuit components appearing in the main part.

\subsection*{Note added}
After the first version of this work appeared on arXiv,
Christoph S\"underhauf kindly informed us that
the method of \cite{SunderhaufCampbellCamps:2023} is actually applicable to our toy problems
and achieves a considerable reduction in necessary resources compared to our naive circuit implementation in Sec.\,\ref{sec:main_analysis}.
In Appendix~\ref{appendix:speedrun}, we present a further-improved implementation inspired by this finding.

\newpage

\section{Review of qubitization of sparse matrices}\label{sec:qubitization_review}
The problem of interest here is to compute the eigenvalues $\lambda$ of a given $N\times N$ Hermitian matrix $H$.
In order to feed the matrix $H$ to a quantum computer, one has to somehow embed it in a unitary matrix $U_H$.
One naive embedding is just to take $U_H=e^{icH}$ where $c$ is some constant.
Although this is conceptually simple,
actual implementation as a quantum circuit is often costly, making its practical use prohibitive.

There is another embedding called \emph{block encoding}
where $U_H$ is an $(N+\Delta N)\times (N+\Delta N)$ unitary matrix
and is reduced to (suitably rescaled) $H$ using a certain \textit{state} vector $\ket{\psi}$ of size $\Delta N$ as
\begin{equation*}
	\braket{\psi|\,U_H\,|\psi} = H.
\end{equation*}
This type of embedding is obviously not unique, and it turns out \cite{LowChuang:2016} that
among them is an especially nice one $U_H^{\ast}$ such that
\begin{equation}\label{eq:eigenvector_W}
	U_H^{\ast}\bigg(
		\frac{
			\ket{\lambda}\ket{\psi} \pm i\ket{\text{orthogonal}}
		}{\sqrt{2}}
	\bigg)
	=
	e^{\mp i\arccos{\lambda}} \bigg(
		\frac{
			\ket{\lambda}\ket{\psi} \pm i\ket{\text{orthogonal}}
		}{\sqrt{2}}
	\bigg)
\end{equation}
where $H\ket{\lambda} = \lambda\ket{\lambda}$ and $(\bra{\psi}\bra{\lambda})\ket{\text{orthogonal}} = 0$.
\emph{Qubitization} 
is a method to generate $U_H^{\ast}$ from a block encoding $U_H$,
and by applying QPE to $U_H^\ast$, one can extract the desired information on $\lambda$
with far fewer computational costs compared to the naive embedding.

Fortunately,
systematic ways of (efficient) block encoding are known
for several types of matrices with special properties or structures.
Previous studies on cost estimation have mostly taken up those with representations in a
linear combination of unitaries,
appearing in quantum chemistry/physics problems e.g. as a result of Jordan-Wigner transformation of underlying fermionic Hamiltonians.
However, there is another nice class of matrices with such systematic block encoding,
namely \emph{$d$-sparse} matrices\footnote{
	Here we follow \cite{LowChuang:2016} and adopt the symbol $d$ (presumably after \emph{density}).
} with at most $d$ non-zero entries in each row (and/or column).\footnote{
	Of course one can call any $N\times N$ matrix ``$N$-sparse,''
	but the computational advantage arises only for sparse enough matrices i.e. $d \ll N$.
	(The same is true for the case of LCU;
	any matrix can be written in a form of $\sum_i \alpha_i U_i$,
	but the whole algorithm can be carried out efficiently only for those with small $\|\vec{\alpha}\|$.)
}
In the following, we introduce two special oracles $O_F, O_H$
and review how the block encoding and the qubitization for $d$-sparse matrices are realized using them.

\subsection{Block encoding}\label{subsec:block-encoding}
One way to block-encode an $N\times N$ Hermitian matrix $H$
is by using two oracles $O_F$ and~$O_H$ \cite{BerryChilds2009blackbox, LowChuang:2016} (see also \cite[Lem.\,48]{GilyenSuLowWiebe:2018}) such that
for $x\in\{1,\dots,N\}$ and $i\in\{1,\dots,d\}$,
\begin{equation*}
    O_F \ket{x,i}=\ket{x,y_i}
\end{equation*}
where $y_i \in \{1,\dots,N\}$ is the position (i.e. column index) of the $i$-th non-zero element in the $x$-th row, and
for $x,y\in\{1,\dots,N\}$,
\begin{equation*}
    O_H \ket{x,y}\ket{z}=\ket{x,y}\ket{z\oplus H_{xy}}
\end{equation*}
where ($z$ is any number and) $H_{xy}$ is the $(x,y)$-element of $H$, hereafter assumed to be (real) non-negative,\footnote{
	The restriction is to avoid subtleties regarding signs and phases arising from the square roots in Eq.\,\eqref{eq:phi_psi};
	for a completely general treatment, see \cite[Sec.\,III]{BerryChilds2009blackbox}.
	This indeed holds for the examples in Sec.\,\ref{sec:main_analysis}.
}
and $\oplus$ denotes a bitwise XOR.

Let
$a_1, a_2$ denote ancillary single-qubit registers and
$a_s, s$ denote ancillary and signal $\lceil\log_2 N\rceil$-qubit registers respectively. 
Our initial target is a unitary operator
\begin{equation*}
	U_H = U_2^\dagger U_1
\end{equation*}
where two unitaries $U_1, U_2$ are such that
\begin{equation}\label{eq:phi_psi}
	\begin{array}{lcl}
		U_1 \ket{0}_{a_1}\ket{0}_{a_2}\ket{0}_{a_s}\ket{x'}_s
		& = &
		\displaystyle
		\frac{1}{\sqrt{d}}\sum_{y'} \Big(
			\sqrt{\frac{H_{x'y'}}{\smash{\underset{i,j}{\max}\,H_{ij}}}}\ket{0}_{a_1} +
			\sqrt{1-\frac{H_{x'y'}}{\smash{\underset{i,j}{\max}\,H_{ij}}}}\ket{1}_{a_1}
		\Big)
		\ket{0}_{a_2}\ket{y'}_{a_s}\ket{x'}_s\\[16pt]
		U_2  \ket{0}_{a_1}\ket{0}_{a_2}\ket{0}_{a_s}\ket{x}_s
		& = &
		\displaystyle
		\ket{0}_{a_1}
		\frac{1}{\sqrt{d}}\sum_y \Big(
			\sqrt{\frac{H_{xy}}{\smash{\underset{i,j}{\max}\,H_{ij}}}}\ket{0}_{a_2} +
			\sqrt{1-\frac{H_{xy}}{\smash{\underset{i,j}{\max}\,H_{ij}}}}\ket{1}_{a_2}
		\Big)
		\ket{x}_{a_s}\ket{y}_s
	\end{array}
\end{equation}
and the summations are over $y$ (resp. $y'$) whose corresponding matrix elements $H_{xy}$ (resp.~$H_{x'y'}$) are non-zero.
This unitary $U_H$ block-encodes the original matrix $H$ as
\begin{equation*}
	\bra{0}_{a_1}\bra{0}_{a_2}\bra{0}_{a_s}\bra{x}_{s} U_H
	\ket{0}_{a_1}\ket{0}_{a_2}\ket{0}_{a_s} \ket{x'}_{s}
	=
	\frac{1}{d \cdot \underset{i,j}{\max}\,H_{ij}}
	\underbrace{
		\sum_{y}\sum_{y'} \sqrt{H_{xy} H_{x'y'}} \braket{x|y'}\braket{y|x'}
	}_{
		= H_{xx'}
	}
\end{equation*}
i.e.
$\bra{0}_{a_1}\bra{0}_{a_2}\bra{0}_{a_s} U_H
\ket{0}_{a_1}\ket{0}_{a_2}\ket{0}_{a_s}
\sim
H$.

\newpage

Starting from a state $\ket{0}_{a_1}\ket{0}_{a_2}\ket{0}_{a_s}\ket{x'}_s\ket{0}_v$ with an additional register $v$ of suitable size,
one can easily verify that $U_1$ can be realized for example by sequentially acting
\begin{enumerate}
	\item \emph{diffusion} gates (e.g. Hadamard gates $H^{\otimes \log_2 d}$ when $d$ is a power of two),\\
	making an equal-superposition state $\displaystyle \frac{1}{\sqrt{d}}\sum_{i=1}^d \ket{0}_{a_1}\ket{0}_{a_2}\ket{x',i}_{s,a_s}\ket{0}_v$,\\[-14pt]
	\item $O_F$, turning the state into $\displaystyle \frac{1}{\sqrt{d}}\sum_{i=1}^d \ket{0}_{a_1}\ket{0}_{a_2}\ket{x',y'_i}_{s,a_s}\ket{0}_v$,\\[-14pt]
	\item $O_H$, loading the values of matrix elements as
	$\displaystyle \frac{1}{\sqrt{d}}\sum_{i=1}^d \ket{0}_{a_1}\ket{0}_{a_2}\ket{x',y'_i}_{s,a_s}\ket{H_{x'y'_i}}_v$,
	\item a controlled rotation,
	leading to\\[2pt]
	$\displaystyle \frac{1}{\sqrt{d}}\sum_{y'} \Big(
		\sqrt{\frac{H_{x'y'}}{\smash{\underset{i,j}{\max}\,H_{ij}}}}\ket{0}_{a_1} +
		\sqrt{1-\frac{H_{x'y'}}{\smash{\underset{i,j}{\max}\,H_{ij}}}}\ket{1}_{a_1}
	\Big)\ket{0}_{a_2}\ket{x',y'}_{s,a_s}\ket{H_{x'y'}}_v$,
	\item $O_H$ again, uncomputing the state and returning the $v$ register to $\ket{0}_v$
	(as $k \oplus k = 0$ holds for any integer $k$), and thus achieving the desired state as in Eq.\,\eqref{eq:phi_psi}.
\end{enumerate}
Note that the controlled rotation can actually be done \emph{in-place}, making the $v$ register unnecessary and simplifying Step $3$\,--\,$5$.
Similarly, starting from a state $\ket{0}_{a_1}\ket{0}_{a_2}\ket{0}_{a_s}\ket{x}_s\ket{0}_v$,
one can check that $U_2$ can be realized by sequentially acting
\begin{enumerate}
	\item diffusion gates, making an equal-superposition state $\displaystyle \frac{1}{\sqrt{d}}\sum_{i=1}^d \ket{0}_{a_1}\ket{0}_{a_2}\ket{x,i}_{s,a_s}\ket{0}_v$,\\[-14pt]
	\item $O_F$, turning the state into $\displaystyle \frac{1}{\sqrt{d}}\sum_{i=1}^d \ket{0}_{a_1}\ket{0}_{a_2}\ket{x,y_i}_{s,a_s}\ket{0}_v$,\\[-14pt]
	\item $O_H$, loading the values of matrix elements as
	$\displaystyle \frac{1}{\sqrt{d}}\sum_{i=1}^d \ket{0}_{a_1}\ket{0}_{a_2}\ket{x,y_i}_{s,a_s}\ket{H_{xy_i}}_v$,
	\item a controlled rotation, leading to \\[2pt]
	$\displaystyle \ket{0}_{a_1}\frac{1}{\sqrt{d}}\sum_y \Big(
		\sqrt{\frac{H_{xy}}{\smash{\underset{i,j}{\max}\,H_{ij}}}}\ket{0}_{a_2} +
		\sqrt{1-\frac{H_{xy}}{\smash{\underset{i,j}{\max}\,H_{ij}}}}\ket{1}_{a_2}
	\Big)\ket{x,y}_{s,a_s}\ket{H_{xy}}_v$,
	\item $O_H$ again, uncomputing the state,
	\item SWAP gates swapping $a_s$ and $s$ registers, achieving the desired state as in Eq.\,\eqref{eq:phi_psi},
\end{enumerate}
the only difference to $U_1$ being (which qubit to rotate and) the additional swaps at the end.

\newpage

\subsection{Qubitization}\label{subsec:qubitization}
With the unitary $U_H$ thus constructed at hand, one can generate a nice block encoding $U_H^\ast$ called the \emph{walk} or \emph{iterate} operator.
Following \cite[Lem.\,10]{LowChuang:2016},
let us add another qubit and consider a unitary $\tilde{U} \coloneqq \ket{0}\!\bra{0}\otimes U_H + \ket{1}\!\bra{1}\otimes U_H^\dagger$.
Together with a NOT operation on the additional qubit $X \coloneqq (\ket{0}\!\bra{1} + \ket{1}\!\bra{0}) \otimes I_a \otimes I_s$,
the unitary $U_H^\ast$ is defined to be
\begin{equation*}
	U_H^\ast
	\coloneqq
	\Big(
		2\ket{+}\ket{0}_{a_1}\ket{0}_{a_2}\ket{0}_{a_s}
		\bra{+}\bra{0}_{a_1}\bra{0}_{a_2}\bra{0}_{a_s} \otimes I_s - I \otimes I_a \otimes I_s
	\Big) X\tilde{U}
\end{equation*}
which can be realized as
\begin{equation*}
	\begin{tikzpicture}
		\node[scale=1] at (0,0) {
			\begin{quantikz}[row sep={20pt,between origins}, column sep=10pt]
				\lstick{$\ket{1}_{\mathrm{sgn}}$} & \qw & \qw & \qw & \qw & \qw & \gate{Z} & \gate{Z} & \qw\\
				& \qw & \octrl{1} & \ctrl{1} & \targ{} & \gate{H} & \octrl{-1} & \gate{H} & \qw\\
				\lstick{$a_1$} & \qw
				& \gate[wires=8, nwires={4,7}]{U_H}
				& \gate[wires=8, nwires={4,7}]{U_H^\dagger} & \qw & \qw & \octrl{-1}& \qw & \qw\\
				\lstick{$a_2$} & \qw & \qw & \qw & \qw & \qw & \octrl{-1} & \qw & \qw\\
				\lstick[wires=3]{$a_s$} & \qw & \qw & \qw & \qw & \qw & \octrl{-1} & \qw & \qw\\
				& \vdots &&&&& \vdots & \vdots &\\
				& \qw & \qw & \qw & \qw & \qw & \octrl{-1} & \qw & \qw\\
				\lstick[3]{$s$} & \qw & \qw & \qw & \qw & \qw & \qw & \qw & \qw\\
				& \vdots &&&&&& \vdots &\\
				& \qw & \qw & \qw & \qw & \qw & \qw & \qw & \qw
			\end{quantikz}
		};
	\end{tikzpicture}
\end{equation*}
with yet another qubit (at the top in the circuit diagram) identifying a potential sign flip resulting from the reflection operation.
Recalling Eq.\,\eqref{eq:eigenvector_W}, one can then carry out an ordinary QPE against this unitary to compute the phase $\arccos \lambda$,
from which the desired eigenvalue $\lambda$ of the original Hermitian matrix $H$ can be immediately extracted.

\newpage

\section{Application to modal analysis}\label{sec:main_analysis}
Armed with the theory of qubitization, let us see how it is actually worked out.
The target is a system of coupled oscillators with which we are very familiar as a basic model of various phenomena in Nature.
As will be shown below, the analyses often involve solving eigenvalue problems of sparse matrices,
and thus is within a scope of the method introduced in the previous section.
For the sake of simplicity, we always take the matrix size to be $N=2^n$.

\subsection{Settings}
We will be considering a system of $N=2^n$ point masses, linearly (i.e. in one dimension) and periodically connected by springs:
\begin{center}
	\begin{tikzpicture}
		\node at (-4.5,0) {$\cdots$};
		\draw (-4.0,0) -- (-3.8,0);
		\draw (-3.3,0) circle [radius=0.5]  node[above=0.6]{$m_{i-1}$};
		\draw(-2.8,0) -- (-2.6,0);
		\draw[
			snake=coil,
			segment amplitude=8,
			segment length=4
		] (-2.6,0) -- node[below=0.5] {$k_{i-1}$} (-0.6,0);
		\draw (-0.6,0) -- (-0.5,0);
		\draw (0,0) circle [radius=0.5] node[above=0.6]{$m_i$};
		\draw(0.5,0) -- (0.7,0);
		\draw[
			snake=coil,
			segment amplitude=8,
			segment length=4
		] (0.7,0) -- node[below=0.5] {$k_i$} (2.7,0);
		\draw (2.7,0) -- (2.8,0);
		\draw (3.3,0) circle [radius=0.5] node[above=0.6]{$m_{i+1}$};
		\draw(3.8,0) -- (4.0,0);
		\draw[
			snake=coil,
			segment amplitude=8,
			segment length=4
		] (4.0,0) -- node[below=0.5] {$k_{i+1}$}  (6,0);
		\node at (6.5,0) {$\cdots$};
		\draw[line width=1.5pt, ->] (-5.3,-1.66) -- (7.3,-1.66) node[right] {position};
		\draw (0,-1.51) -- (0,-1.81) node[below=0.2] {$x_i$};
		\draw (3.3,-1.51) -- (3.3,-1.81) node[below=0.2] {$x_{i+1}$};
		\draw (-3.3,-1.51) -- (-3.3,-1.81) node[below=0.2] {$x_{i-1}$};
	\end{tikzpicture}
\end{center}
Here, $m_i$'s denote masses and $k_i$'s denote spring constants.
The equation of motion for the $i$-th mass reads
\begin{equation*}
    m_i \ddot{x}_i=k_i(x_{i+1}-x_i)-k_{i-1}(x_i-x_{i-1})
\end{equation*}
(with $x_{0}\coloneqq x_{N}$ and $x_{N+1}\coloneqq x_1$ etc.),
and the natural frequencies (resp. normal modes) are given as eigenvalues (resp. eigenvectors) of an $N\times N$ matrix
\begin{equation}\label{eq:original-matrix}
	H^{\text{original}} = 
	\left(
		\begin{array}{cccccc}
			-\frac{k_N+k_1}{m_1} & \frac{k_1}{m_1} & 0 & \cdots & 0 & \frac{k_N}{m_1}\\[2pt]
			\frac{k_1}{m_2} & -\frac{k_1+k_2}{m_2} & \frac{k_2}{m_2} & 0 & \cdots & 0\\[2pt]
			0 & \frac{k_2}{m_3} & -\frac{k_2+k_3}{m_3} &&& \vdots\\[2pt]
			\vdots &&& \rotatebox[]{-10}{$\ddots$} && 0\\[2pt]
			0 &&&& -\frac{k_{N-2}+k_{N-1}}{m_{N-1}} & \frac{k_{N-1}}{m_{N-1}}\\[2pt]
			\frac{k_N}{m_N} & 0 & \cdots & 0 & \frac{k_{N-1}}{m_N} & -\frac{k_{N-1}+k_N}{m_N}
		\end{array}
	\right).
\end{equation}

\newpage

For the toy problems examined in this note,
we restrict ourselves to matrices \eqref{eq:original-matrix} whose diagonal elements are all equal.
This allows us to completely separate the diagonal part and the off-diagonal part and to forget about the former.
As a result, target matrices $H$'s we want to diagonalize are 2-sparse (i.e. $d=2$) with very simple structure, namely
\begin{equation}\label{eq:2-sparse-matrix}
	H_{xy} 
	\left\{
		\begin{array}{cl}
			\neq 0 & (|x-y| =1 \text{ mod }2^n),\\
			= 0 & (\text{otherwise}).\\
		\end{array}
	\right.
\end{equation}
Note that these simplicities are just for demonstration;
our methods can be easily generalized to treat matrices with e.g.
\begin{itemize}
	\item non-identical diagonal elements
	\item non-periodic boundary conditions
	\item wider ``band'' (which in particular incorporate $k\,(\geq 2)$-dimensional cases with more complicated couplings between multiple coordinates described by $kN\times kN$ matrices)
\end{itemize}
as will be described along the way.

\subsection{Adder}
One of the fundamental components in implementation of the oracles $O_F, O_H$ is an adder.
Given two states $\ket{x}, \ket{y}$ encoding (binary representations of) integers $x,y$ modulo $2^n$ in a computational basis,
the (required) function of the adder is to return a state $\ket{x+y}$ encoding $(x+y)$ modulo $2^n$.
In this note, we adopt the adder proposed by \cite{CuccaroDraperKutinMoulton:2004},
whose I/O is schematically represented as follows
\begin{equation*}
	\begin{tikzpicture}
		\node[scale=1] at (0,0) {
			\begin{quantikz}[row sep={20pt,between origins}, column sep=10pt]
				\lstick[wires=3]{$\ket{x}_x$} \qw & \gate[wires=6, nwires={2,5}]{\ \mathrm{adder}\ } & \qw \rstick[wires=3]{$\ket{x}_x$}\\
				\vdots && \vdots \\
				\qw && \qw \\
				\lstick[wires=3]{$\ket{y}_y$} \qw && \qw \rstick[wires=3]{$\ket{x+y}_y$}\\
				\vdots && \vdots \\
				\qw && \qw 
			\end{quantikz}
		};
	\end{tikzpicture}
\end{equation*}
and consumes $2n$ Toffoli gates (plus a single ancillary qubit which is omitted above).
For the details of implementation, see Appendix~\ref{subsec:Adder}.

\subsection{Oracle $O_F$}\label{subsec:O_F}
For the matrices \eqref{eq:2-sparse-matrix}, non-trivial computations involving the oracle $O_F$ are limited to\footnote{
	The index origin is (shifted by one from the expressions in Sec.\,\ref{sec:qubitization_review} and) set to zero for convenience.
}
\begin{align*}
    O_F\ket{x,0}&=\ket{x,x-1},\\
    O_F\ket{x,1}&=\ket{x,x+1}.
\end{align*}
Recalling that flipping all bits of an integer $y$ leads to 
$(2^n-1)-y \equiv (-y-1)$ modulo $2^n$,
this $O_F$ can be implemented in a rather straightforward manner
employing suitable ancillary qubits encoding ``$\ket{-1}$'' as 
\begin{equation*}
	\begin{tikzpicture}
		\node[scale=1] at (-8,0) {
			\begin{quantikz}[row sep={20pt,between origins}, column sep=10pt]
				\lstick[wires=3]{$\ket{x}_x$} \qw & \gate[wires=10, nwires={2,5,6,9}]{O_F} & \qw \rstick[wires=3]{$\ket{x}_x$}\\
				\vdots && \vdots \\
				\qw && \qw \\
				\lstick[wires=4]{$\ket{i}_y$} \qw && \qw \rstick[wires=4]{$\ket{x\pm 1}_y$}\\
				\vdots && \vdots \\
				 && \\
				\qw && \qw \\
				\lstick[wires=3]{$\ket{0}_a$} \qw && \qw \rstick[wires=3]{$\ket{0}_a$}\\
				\vdots && \vdots \\
				\qw && \qw
			\end{quantikz}
		};
		\node at (-4.7,0) {$=$};
		\node[scale=1] at (0,0) {
			\begin{quantikz}[row sep={20pt,between origins}, column sep=10pt]
				\qw & \qw
				& \qw
				& \gate[wires=7, nwires={2,5}]{\ \ \ \mathrm{adder}\ \ \ } & \qw & \qw & \qw\\
				& \vdots &&&& \vdots &\\
				& \qw & \qw & \qw & \qw & \qw & \qw\\
				\qw & \qw & \qw & \gateoutput[4]{$\ket{x+2i}$} &  \gate[wires=7, nwires={2,6}]{\ \ \ \mathrm{adder}\ \ \ }\gateoutput[4]{$\ket{x+2i-1}$} & \qw & \qw\\
				& \vdots &&&& \vdots &\\
				& \targ{} & \ctrl{1} & \qw & \qw & \qw & \qw \\
				& \ctrl{-1} & \targ{} & \qw & \qw & \qw & \qw \\
				\qw & \targ{} & \qw & \qw & \qw & \targ{} & \qw\\
				& \vdots &&&& \vdots &\\
				& \targ{} & \qw & \qw & \qw & \targ{} & \qw
			\end{quantikz}
		};
	\end{tikzpicture}
\end{equation*}
with each register $x,y,a$ consisting of $n$ qubits.
Here, the two CNOT gates in front of the first adder \emph{doubles} the input $i$,
mapping $\ket{i=0}_y \mapsto \ket{0}_y$ and $\ket{i=1}_y \mapsto \ket{2}_y$, respectively.\footnote{
	In general, one can use another qubit to first somehow check whether $0 \leq i\leq d-1$ actually holds and output the result to the qubit,
	and then apply the gates additionally controlled on the qubit, if necessary.
	For small enough $d$, brute-force multi-controlled NOT gates would do the job without ruining efficiency.
}
Also, note that the results of the addition are always output to the $y$ register,
meaning in particular that the second adder is placed ``upside down.''

Since it contains two adders of $n$-bit integers,
the circuit as a whole consumes $2 \times 2n = 4n$ Toffoli gates with $(3n+1)$ qubits,
where the second adder recycles the ancillary qubit of the first adder (which is omitted in the diagram).

\newpage

\subsection{Oracle $O_H$}\label{subsec:O_H}
When non-zero $H_{xy}$'s of the matrices \eqref{eq:2-sparse-matrix} have simple dependence on a difference $(x-y)$,
a naive way to implement the oracle computation $O_H\ket{x,y}\ket{z} = \ket{x,y}\ket{z\oplus H_{xy}}$
is to first read out $(x-y)$ and then bitwise-XOR the corresponding element $H_{xy}$.
The former can be realized for example by adding $x$ and $(-y-1)$,
and the latter can be realized by (hard-coded) bitwise NOT gates controlled on the difference,
i.e. by placing NOT gates acting on qubits corresponding to \textit{set-bits} of the number $H_{xy}$.
The quantum circuit is as
\begin{equation*}
	\begin{tikzpicture}
		\node[scale=0.9] at (0,0) {
			\begin{quantikz}[row sep={20pt,between origins}, column sep=10pt]
				\lstick[wires=3]{$\ket{x}_x$} \qw & \qw & \gate[wires=7, nwires={2,5}]{\ \mathrm{adder}\ } & \qw & \qw & \qw & \qw & \qw \rstick[wires=3]{$\ket{x}_x$}\\
				& \vdots &&&&& \vdots &\\
				& \qw{} & \qw & \qw & \qw & \qw & \qw & \qw\\
				\lstick[wires=4]{$\ket{y}_y$} \qw & \targ{} & \qw & \qw & \octrl{1} & \ctrl{1} & \qw & \qw\rstick[wires=4]{$\ket{x-y-1}_y$}\\
				& \vdots &&& \vdots & \vdots & \vdots &\\
				& \targ{} & \qw & \qw & \octrl{1} & \ctrl{1} & \qw & \qw\\
				& \targ{} & \qw & \qw & \octrl{1} & \octrl{1} & \qw & \qw\\
				\lstick[wires=3]{$\ket{z}_z$} \qw & \qw & \qw & \qw
				& \gate[wires=3, nwires=2]{\oplus H_{x(x-1)}}
				& \gate[wires=3, nwires=2]{\oplus H_{x(x+1)}} & \qw & \qw\rstick[wires=3]{$\ket{z \oplus H_{xy}}_z$}\\
				& \vdots &&&&& \vdots &\\
				& \qw & \qw & \qw & \qw & \qw \ \cdots\ & \qw & \qw
			\end{quantikz}
		};
	\end{tikzpicture}
\end{equation*}
followed by suitable uncomputation, namely
\begin{equation*}
	\begin{tikzpicture}
		\node[scale=1] at (-8.66,0) {
			\begin{quantikz}[row sep={20pt,between origins}, column sep=10pt]
				\lstick[wires=3]{$\ket{x}_x$} \qw & \gate[wires=10, nwires={2,5,6,9}]{O_H} & \qw \rstick[wires=3]{$\ket{x}_x$}\\
				\vdots && \vdots \\
				\qw && \qw \\
				\lstick[wires=4]{$\ket{y}_y$} \qw && \qw \rstick[wires=4]{$\ket{y}_y$}\\
				\vdots && \vdots \\
		 		&& \\
				\qw && \qw \\
				\lstick[wires=3]{$\ket{z}_z$} \qw && \qw \rstick[wires=3]{$\ket{z\oplus H_{xy}}_z$}\\
				\vdots && \vdots \\
				\qw && \qw
			\end{quantikz}
		};
		\node at (-6.6,0) {$=$};
		\node[scale=1] at (0,0) {
			\begin{quantikz}[row sep={20pt,between origins}, column sep=10pt]
				\qw & \qw & \gate[wires=7, nwires={2,5}]{\ \mathrm{adder}\ } & \qw & \qw & \qw & \qw & \gate[wires=7, nwires={2,5}]{\ \mathrm{adder}^\dagger\ } & \qw  & \qw \\
				& \vdots &&&&&&& \vdots &&\\
				& \qw{} & \qw & \qw & \qw & \qw & \qw & \qw & \qw & \qw \\
				\qw & \targ{} & \qw & \qw & \octrl{1} & \ctrl{1} & \qw & \qw & \targ{}  & \qw \\
				& \vdots &&& \vdots &&&& \vdots &\\
				& \targ{} & \qw & \qw & \octrl{1} & \ctrl{1} & \qw & \qw& \targ{}  & \qw\\
				& \targ{} & \qw & \qw & \octrl{1} & \octrl{1}  & \qw & \qw & \targ{}  & \qw\\
				\qw & \qw & \qw & \qw
				& \gate[wires=3, nwires=2]{\oplus H_{x(x-1)}}
				& \gate[wires=3, nwires=2]{\oplus H_{x(x+1)}} & \qw & \qw & \qw & \qw\\
				& \vdots &&&&&&& \vdots &\\
				& \qw & \qw & \qw & \qw & \qw\ \cdots\ & \qw & \qw & \qw & \qw
			\end{quantikz}
		};
	\end{tikzpicture}
\end{equation*}
again with $n$-qubit registers $x,y$ and a register $z$ with suitable size to store values.\footnote{
	For more generic band matrices, one can suitably add corresponding multi-controlled bitwise-XOR gates for $H_{x(x\pm 2)}$, $H_{x(x\pm 3)}$, and so on
	(where there is also a chance that, by sorting multi-controlled operations, adjacent ones can be contracted).
}
This time the circuit consumes $2 \times 2n + 2\times 2(n-1) = 8n-2$ Toffoli gates by a naive counting,
as the multi-controlled gates are decomposed as in Appendix~\ref{subsec:Multi-controlled} with $2(n-1)$ Toffoli gates.\footnote{
	If the output of the first adder is guaranteed to be either $+1$ or $-1$,
	one does not need to multi-control the $\oplus H_{xy}$ gate and a single-control operation is sufficient.
	This reduces the number of Toffoli gates used in the circuit and thus makes the computation faster
	(and alludes to the yet-more-efficient implementation presented in Appendix \ref{appendix:speedrun}),
	but here we stick to the naive implementation for ease of understanding.
}

As already mentioned in Sec.\,\ref{sec:qubitization_review},
for actual usage, there is no need to first load the value of $H_{xy}$ and then do the corresponding rotation;
one can directly implement the rotations in-place. 
The quantum circuit of the modified oracle is given as
\begin{equation*}
	\begin{tikzpicture}
		\node[scale=0.8] at (-9,0.1) {
			\begin{quantikz}[row sep={20pt,between origins}, column sep=10pt]
				\lstick[wires=3]{$\ket{x}_x$} \qw & \gate[wires=8, nwires={2,5,6}]{O_H^{\text{mod.}}} & \qw \rstick[wires=3]{$\ket{x}_x$}\\
				\vdots && \vdots \\
				\qw && \qw \\
				\lstick[wires=4]{$\ket{y}_y$} \qw && \qw \rstick[wires=4]{$\ket{y}_y$}\\
				\vdots && \vdots \\
		 		&& \\
				\qw && \qw \\[10pt]
				\lstick{$\ket{0}_z$} \qw && \qw
			\end{quantikz}
		};
		\node at (-7,0) {$=$};
		\node[scale=0.8] at (0,0) {
			\begin{quantikz}[row sep={20pt,between origins}, column sep=10pt]
				\qw & \qw & \gate[wires=7, nwires={2,5}]{\ \ \mathrm{adder}\ \ } & \qw & \qw & \qw & \gate[wires=7, nwires={2,5}]{\ \mathrm{adder}^\dagger\ } & \qw & \qw \\
				& \vdots &&&&&& \vdots \\
				& \qw & \qw & \qw & \qw & \qw & \qw & \qw & \qw\\
				\qw & \targ{} & \qw & \qw & \octrl{1} & \ctrl{1} & \qw & \targ{} & \qw \\
				& \vdots &&& \vdots & \vdots && \vdots &&\\
				& \targ{} & \qw & \qw & \octrl{1} & \ctrl{1} & \qw & \targ{} & \qw\\
				& \targ{} & \qw & \qw & \octrl{1} & \octrl{1} & \qw & \targ{} & \qw\\[10pt]
				\qw & \qw & \qw & \qw
				& \gate{R_y(2\arccos\sqrt{H_{x(x-1)}})}
				& \gate{R_y(2\arccos\sqrt{H_{x(x+1)}})} & \qw & \qw & \qw 
			\end{quantikz}
		};
	\end{tikzpicture}
\end{equation*}

Another point worth mentioning here is about exceptional operations including those concerning ``boundary conditions.''
In Eq.\,\eqref{eq:2-sparse-matrix}, the condition for $H_{xy}$ to be (possibly) non-zero involved taking modulo $2^n$ (due to the periodic boundary condition of the system),
resulting in a \emph{circulant} matrix.
However, by inserting a bitwise-XOR (or corresponding rotation) gate controlled on the $x$ register in addition to the $y$ register
between two adders in the above circuit, one can selectively \emph{undo} the operation and kill unnecessary elements.
For example, by acting controlled-$R_y(-2\arccos\sqrt{H_{x(x-1)}})$ gate only when $x=0$,
one can in effect eliminate the $H_{0(n-1)}$ element.
Doing the same thing for $H_{(n-1)0}$, one can realize an oracle $O_H$ for
a \emph{tridiagonal} (or more generally a \emph{band}) version of the original matrix (corresponding to a fixed boundary condition), without much cost.

\setlength{\footnotemargin}{14pt}

\subsection{Example 1: equal spring constants}\label{subsec:ex1}
Now let us consider a simple system of $N=2^n$ point masses of equal mass $m$ connected by springs with equal spring constant $k$.
The matrix \eqref{eq:original-matrix} (up to a constant factor $k/m$) becomes
\begin{equation}\label{eq:Hamiltonian_ex1}
	H_{xy}^{\text{original}} = 
	\left\{
		\begin{array}{cl}
			1 & (|x-y| = 1)\\
			-2 & (|x-y| = 0)\\
			0 & (\text{otherwise})\\
		\end{array}
	\right.
\end{equation}
and one can forget about the diagonal elements by considering $H = H^{\text{original}}+2I$.\footnote{
	The problem itself is just a special case of that considered in \cite[Sec.\,4.2]{CampsLinBeeumenYang:2022},
	but the emphasis is put on somewhat complementary aspects, so to speak. 
}

Since all the non-zero matrix elements take the same value, things are extremely simplified;
the controlled rotation parts of the algorithm become trivial,
and therefore one can safely skip the steps involving ancillary qubits $a_1$, $a_2$, and $v$
(and thus completely discard them),
making the oracle $O_H$ unnecessary as a result.
This indeed makes sense because all we need is mere information about the position of non-zero matrix elements and not the values themselves.

Anyway, following the description in Sec.\,\ref{subsec:block-encoding},
an explicit circuit for the unitary $U_H$ is given by concatenating two circuit components corresponding to $U_1$ and $U_2^\dagger$
\begin{equation*}
	\begin{tikzpicture}
		\node[scale=0.9] at (0,0) {
			\begin{quantikz}[row sep={20pt,between origins}, column sep=10pt]
				\lstick[wires=4]{$\ket{x'}_s$}
				& \qw & 
				\gate[wires=11, nwires={2,6,10}]{\ \ \ O_F\ \ \ }\gateinput[4]{$x$} & \qw & \qw \rstick[wires=4]{$\ket{x'}_s$}\\
				& \vdots && \vdots &\\
				& \qw & \qw & \qw  & \qw\\
				& \qw & \qw & \qw  & \qw\\
				\lstick[wires=4]{$\ket{0}_{a_s}$}
				& \qw & \gateinput[4]{$i$} & \qw & \qw \rstick[wires=4]{$\ket{y'}_{a_s}$}\\
				& \vdots && \vdots &\\
				& \qw & \qw & \qw & \qw \\
				& \gate{H} &  & \qw & \qw\\[6pt]
				\lstick[wires=3]{$\ket{0}$}
				& \qw && \qw & \qw \rstick[wires=3]{$\ket{0}$}\\
				& \vdots && \vdots &\\
				& \qw & \qw & \qw & \qw
			\end{quantikz}
		};
		\node[scale=0.9] at (7,0) {
			\begin{quantikz}[row sep={20pt,between origins}, column sep=10pt]
				\lstick[wires=4]{$\ket{y}_s$}
				& \qw & 
				\gate[wires=8, nwires={2,6}]{\text{swap}} &
				\gate[wires=11, nwires={2,6,10}]{\ \ \ O_F^\dagger\ \ \ }\gateoutput[4]{$x$}
				& \qw & \qw \rstick[wires=4]{$\ket{x}_s$}\\
				& \vdots &&& \vdots &\\
				& \qw & \qw & \qw & \qw & \qw\\
				& \qw & \qw & \qw & \qw & \qw\\
				\lstick[wires=4]{$\ket{x}_{a_s}$}
				& \qw & \qw & \gateoutput[4]{$i$} & \qw & \qw \rstick[wires=4]{$\ket{0}_{a_s}$}\\
				& \vdots &&&\vdots &\\
				& \qw & \qw & \qw & \qw & \qw\\
				& \qw & \qw & \qw & \gate{H} & \qw\\[6pt]
				\lstick[wires=3]{$\ket{0}$}
				& \qw & \qw & \qw & \qw & \qw \rstick[wires=3]{$\ket{0}$}\\
				& \vdots &&& \vdots &\\
				& \qw & \qw & \qw & \qw & \qw
			\end{quantikz}
		};
	\end{tikzpicture}
\end{equation*}
from which one can immediately obtain a circuit for the unitary $U_H^\ast$ as described in Sec.\,\ref{subsec:qubitization}.

\newpage

An important point to note is that
addition of control qubits to $U_H$ is equivalent to that to the ``swap'' gate alone,
since the net output of the circuit is the same even if the $O_F$-$O_F^\dagger$ pair and the $H$-$H$ pair are not controlled on the additional qubits.\footnote{
	In fact, this is a special case of the trick mentioned in Appendix \ref{subsec:Controlled-U}.
}
Taking this trick into account, the unitary $U_H^\ast$ is reduced to a $(4n+2)$-qubit circuit
(including $n$ ancillary qubits for multi-controlled gates (one of which is also used for adders) omitted in the diagram)
\begin{equation*}
	\begin{tikzpicture}
		\node[scale=0.9] at (0,0) {
			\begin{quantikz}[row sep={20pt,between origins}, column sep=10pt]
				\lstick{$\ket{1}_{\text{sgn}}$} & \qw & \qw & \qw & \qw & \qw & \qw & \qw & \gate{Z} & \gate{Z} & \qw\\
				& \qw & \qw & \qw & \qw & \qw & \targ{} & \gate{H} & \octrl{-1} & \gate{H} & \qw\\[6pt]
				\lstick[wires=4]{$a_s$}
				& \qw & 
				\gate[wires=10, nwires={2,6,9}]{\ \ \ O_F\ \ \ }\gateinput[4]{$i$} & 
				\gate[wires=7, nwires={2,6}]{\text{swap}} &
				\gate[wires=10, nwires={2,6,9}]{\ \ \ O_F^\dagger\ \ \ }\gateoutput[4]{$i$}
				& \qw & \qw & \qw & \octrl{-1} & \qw & \qw\\
				& \vdots &&&& \vdots &&& \vdots & \vdots & \\
				& \qw & \qw & \qw & \qw & \qw & \qw & \qw & \octrl{-1} & \qw & \qw\\
				& \gate{H} & \qw & \qw & \qw & \gate{H} & \qw & \qw & \octrl{-1} & \qw & \qw\\
				\lstick[wires=3]{$s$}
				& \qw & \gateinput[3]{$x$} & \qw & \gateoutput[3]{$x$} & \qw & \qw & \qw & \qw & \qw & \qw\\
				& \vdots &&&& \vdots &&&& \vdots &\\
				& \qw & \qw & \qw & \qw & \qw & \qw & \qw & \qw & \qw & \qw\\[6pt]
				\lstick[wires=3]{$\ket{0}$}
				& \qw & \qw & \qw & \qw & \qw & \qw & \qw & \qw & \qw & \qw\\
				& \vdots &&&& \vdots &\\
				& \qw & \qw & \qw & \qw & \qw & \qw & \qw & \qw & \qw & \qw
			\end{quantikz}
		};
	\end{tikzpicture}
\end{equation*}
where the latter half of (controlled-)$U_H$ and the former half of (controlled-)$U_{H}^\dagger$ which do not involve control qubits are canceled out,
and the two controlled-``swap''s are combined into a single (non-controlled) ``swap.''
As a result, a naive counting of Toffoli gates consumed by a controlled-$U_H^\ast$ (to be fed into the QPE) as a whole 
(again keeping the trick in mind)
is a sum of
\begin{itemize}
	\item two $O_F$'s: $2 \times 4n$
	\item a controlled-``swap'' (cf. Appendix \ref{subsec:swap}): $n$
	\item a multi-controlled gate at the end (cf. Appendix \ref{subsec:Multi-controlled}): $2[(n+2)-1]$
\end{itemize}
which is equal to $11n+2$.

\newpage

\subsection{Example 2: alternating spring constants}\label{subsec:ex2}

Let us next consider a slight generalization where one has alternating two spring constants $k_1, k_2$.
Without loss of generality, one can assume $k_1 < k_2$ and
the equation of motion for the $i$-th point mass to be
\begin{equation*}
    m {\ddot x}_i
    =
    \left\{
    	\begin{array}{lll}
    		k_1(x_{i+1}-x_i)-k_2(x_i-x_{i-1}) & (i\text{ even}),\\
    		k_2(x_{i+1}-x_i)-k_1(x_i-x_{i-1}) & (i\text{ odd}).\\
    	\end{array}
    \right.
\end{equation*}
The matrix \eqref{eq:original-matrix} (up to a constant factor $\frac{1}{m}$) becomes\footnote{
	This is actually equivalent to
	a Hamiltonian of the Su-Schrieffer-Heeger model \cite{SuSchriefferHeeger1979, SuSchriefferHeeger1980},
	which describes a polyacetylene as a dimer chain consisting of alternating single ($\sigma$) and double ($\pi$) bonds,
	and suggests potential utility of non-LCU-based qubitization even in quantum chemistry/physics problems.
}
\begin{equation}\label{eq:Hamiltonian_ex2}
	H_{xy}^{\text{original}} = 
	\left\{
		\begin{array}{cl}
			k_1 & (\text{``$x$ even and $x-y = +1$'' or ``$x$ odd and $x-y = -1$''})\\
			-(k_1+k_2) & (x-y = 0)\\
			k_2 & (\text{``$x$ odd and $x-y = +1$'' or ``$x$ even and $x-y = -1$''})\\
			0 & (\text{otherwise})
		\end{array}
	\right.
\end{equation}
and one can consider $H = \frac{1}{k_2}[H^{\text{original}} + (k_1+k_2)I]$ as a target matrix.

For this problem, we further modify the $O_H^{\text{mod.}}$ oracle;
we add the topmost additional control qubit to the controlled rotation gates to identify the parity of $x$.
Then, non-trivial rotations are carried out for $H_{xy} = \frac{k_1}{k_2}$ (while for $H_{xy} = \frac{k_2}{k_2}=1$ rotations are trivial as $\arccos 1 = 0$ and thus unnecessary), which can be implemented as
\begin{equation*}
	\begin{tikzpicture}
		\node[scale=0.8] at (-8.4,0.1) {
			\begin{quantikz}[row sep={20pt,between origins}, column sep=10pt]
				\lstick[wires=4]{$\ket{x}_x$} \qw & \gate[wires=9, nwires={2,3,6,7}]{O_H^{\text{mod.}}} & \qw \rstick[wires=4]{$\ket{x}_x$}\\
				\vdots && \vdots \\
				&& \\
				\qw && \qw \\
				\lstick[wires=4]{$\ket{y}_y$} \qw && \qw \rstick[wires=4]{$\ket{y}_y$}\\
				\vdots && \vdots \\
		 		&& \\
				\qw && \qw \\[10pt]
				\lstick{$\ket{0}_z$} \qw && \qw
			\end{quantikz}
		};
		\node at (-6.2,0) {$=$};
		\node[scale=0.8] at (0,0) {
			\begin{quantikz}[row sep={20pt,between origins}, column sep=10pt]
				\qw & \qw & \gate[wires=8, nwires={2,6}]{\ \mathrm{adder}\ } & \qw & \qw & \qw & \gate[wires=8, nwires={2,6}]{\ \mathrm{adder}^\dagger\ } & \qw & \qw \\
				& \vdots &&&&&& \vdots &\\
				& \qw{} & \qw & \qw & \qw & \qw & \qw & \qw & \qw\\
				& \qw{} & \qw & \qw & \octrl{1} & \ctrl{1} & \qw & \qw & \qw\\
				\qw & \targ{} & \gateoutput[4]{$x-y-1$}\qw & \qw & \octrl{1} & \ctrl{1} & \qw & \targ{} & \qw \\
				& \vdots &&& \vdots & \vdots && \vdots &&\\
				& \targ{} & \qw & \qw & \octrl{1} & \ctrl{1} & \qw & \targ{} & \qw\\
				& \targ{} & \qw & \qw & \octrl{1} & \octrl{1} & \qw & \targ{} & \qw\\[10pt]
				\qw & \qw & \qw & \qw
				& \gate{R_y(2\arccos\sqrt{\tfrac{k_1}{k_2}})}
				& \gate{R_y(2\arccos\sqrt{\tfrac{k_1}{k_2}})} & \qw & \qw & \qw 
			\end{quantikz}
		};
	\end{tikzpicture}
\end{equation*}

The circuit of the unitary $U_H$ should be given by concatenating $U_1$ and $U_2^\dagger$ as
\begin{equation*}
	\begin{tikzpicture}
		\node[scale=0.9] at (0,0) {
			\begin{quantikz}[row sep={20pt,between origins}, column sep=10pt, transparent]
				\lstick[wires=3]{$\ket{x}_s$}
				& \qw & 
				\gate[wires=12, nwires={2,5,11}]{\ \ \ O_F\ \ \ }\gateinput[3]{$x$} & 
				\gate[wires=8, nwires={2,5}]{\ \ \ O_H^{\text{mod.}}\ \ } & \qw & \qw \rstick[wires=3]{$\ket{x}_s$}\\
				& \vdots &&& \vdots &\\
				& \qw & \qw & \qw & \qw & \qw\\
				\lstick[wires=4]{$\ket{0}_{a_s}$}
				& \qw & \gateinput[4]{$i$} & \gateinput[4]{$y$} & \qw & \qw \rstick[wires=4]{$\ket{y}_{a_s}$}\\
				& \vdots &&& \vdots &\\
				& \qw & \qw & \qw & \qw & \qw\\
				& \gate{H} & \qw & \qw & \qw & \qw\\[6pt]
				\lstick[]{$\ket{0}_{a_1}$}
				& \qw & \linethrough & \gateinput[]{$z$} & \qw & \qw \rstick[]{$\ket{\ast}_{a_1}$}\\
				\lstick[]{$\ket{0}_{a_2}$}
				& \qw & \linethrough & \qw & \qw & \qw \rstick[]{$\ket{0}_{a_2}$}\\
				\lstick[wires=3]{$\ket{0}$}
				& \qw & \qw & \qw & \qw & \qw \rstick[wires=3]{$\ket{0}$}\\
				& \vdots &&& \vdots &\\
				& \qw & \qw & \qw & \qw & \qw
			\end{quantikz}
		};
		\node[scale=0.9] at (9,0) {
			\begin{quantikz}[row sep={20pt,between origins}, column sep=10pt, transparent]
				\lstick[wires=3]{$\ket{y}_s$} \qw & \qw & 
				\gate[wires=7, nwires={2,5}]{\text{swap}} &
				\gate[wires=9, nwires={2,5}]{\ \ \ O_H^{\text{mod.}\dagger}\ \ } &
				\gate[wires=12, nwires={2,6,11}]{\ \ \ O_F^\dagger\ \ \ }\gateoutput[3]{$x$}
				& \qw & \qw \rstick[wires=3]{$\ket{x}_s$}\\
				& \vdots &&&& \vdots &\\
				\qw & \qw & \qw & \qw & \qw & \qw & \qw\\
				\lstick[wires=4]{$\ket{x}_{a_s}$} \qw & \qw & \qw & \qw & \gateoutput[4]{$i$} & \qw &\qw \rstick[wires=4]{$\ket{0}_{a_s}$}\\
				& \vdots &&&& \vdots &\\
				\qw & \qw & \qw & \qw & \qw & \qw & \qw\\
				\qw & \qw & \qw & \qw & \qw & \gate{H} & \qw\\[6pt]
				\lstick[]{$\ket{0}_{a_1}$} \qw & \qw & \qw & \linethrough & \linethrough & \qw & \qw \rstick[]{$\ket{0}_{a_1}$}\\
				\lstick[]{$\ket{\ast}_{a_2}$} \qw & \qw & \qw & \gateoutput[]{$z$} & \linethrough & \qw & \qw \rstick[]{$\ket{0}_{a_2}$}\\
				\lstick[wires=3]{$\ket{0}$} \qw & \qw & \qw & \qw \qw & \qw & \qw & \qw \rstick[wires=3]{$\ket{0}$}\\
				& \vdots &&&& \vdots &\\
				\qw & \qw & \qw & \qw & \qw & \qw & \qw
			\end{quantikz}
		};
	\end{tikzpicture}
\end{equation*}
As before, when adding control qubits to $U_H$, one does not need to add them to the $O_F$-$O_F^\dagger$ pair, the $H$-$H$ pair,
and the adder-adder$^\dagger$ pairs inside $O_H^{\text{mod.}}$'s.
As a result, $U_H^\ast$ is reduced to a $(4n+6)$-qubit circuit (including $(n+2)$ ancillary qubits (suitably recycled) for the multi-controlled gates) as
\begin{equation*}
	\begin{tikzpicture}
		\node[scale=0.66] at (0,0) {
			\begin{quantikz}[row sep={20pt,between origins}, column sep=10pt, transparent]
				\lstick{$\ket{1}_{\text{sgn}}$} & \qw & \qw & \qw & \qw & \qw & \qw & \qw & \qw & \qw & \qw & \qw & \qw & \qw & \qw & \qw & \qw  & \qw & \qw & \gate{Z} & \gate{Z} & \qw\\
				\lstick{} & \qw & \qw & \octrl{1} & \octrl{1} & \qw & \qw & \octrl{4} & \octrl{4} & \ctrl{4} & \ctrl{4} & \qw & \qw & \ctrl{1} & \ctrl{1} & \qw & \qw & \targ{} & \gate{H} & \octrl{-1} & \gate{H} & \qw\\[6pt]
				\lstick[wires=4]{$s$}
				& \qw & 
				\gate[wires=13, nwires={2,6,12}]{O_F} & 
				\gate[wires=9, nwires={2,6}]{O_H^{\text{mod.}}} &
				\gate[wires=8, nwires={2,6}]{\text{swap}} & \qw &
				\gate[wires=8, nwires={2,6}]{\text{adder}} & \qw & \qw & \qw & \qw &
				\gate[wires=8, nwires={2,6}]{\text{adder}^\dagger} & \qw &
				\gate[wires=8, nwires={2,6}]{\text{swap}} &
				\gate[wires=9, nwires={2,6}]{O_H^{\text{mod.}\dagger}} &
				\gate[wires=13, nwires={2,6,12}]{O_F^\dagger} & \qw & \qw & \qw & \qw & \qw & \qw\\
				& \vdots &&&&&&&&&&&&&&&&&&& \vdots &\\
				& \qw & \qw & \qw & \qw & \qw & \qw & \qw & \qw & \qw & \qw & \qw & \qw & \qw & \qw & \qw & \qw & \qw & \qw & \qw & \qw & \qw\\
				& \qw & \qw & \qw & \qw & \qw & \qw & \octrl{1} & \ctrl{1} & \ctrl{1} & \octrl{1} & \qw & \qw & \qw & \qw & \qw & \qw & \qw & \qw & \qw & \qw & \qw\\
				\lstick[wires=4]{$a_s$}
				& \qw & \qw & \qw & \qw & \targ{} & \qw & \octrl{1} & \ctrl{1} & \ctrl{1} & \octrl{1} & \qw & \targ{} & \qw & \qw & \qw & \qw & \qw & \qw & \octrl{-5} & \qw & \qw\\
				& \vdots &&&& \vdots && \vdots & \vdots & \vdots & \vdots && \vdots &&&& \vdots &&& \vdots & \vdots &\\
				& \qw & \qw & \qw & \qw & \targ{} & \qw & \octrl{1} & \ctrl{1} & \ctrl{1} & \octrl{1} & \qw & \targ{}  & \qw & \qw & \qw & \qw & \qw & \qw & \octrl{-1} & \qw & \qw\\
				& \gate{H} & \qw & \qw & \qw & \targ{} & \qw & \octrl{2} & \octrl{2} & \octrl{2} & \octrl{2} & \qw & \targ{} & \qw & \qw & \qw & \gate{H} & \qw & \qw & \octrl{-1} & \qw & \qw\\[6pt]
				\lstick[]{$a_1$}
				& \qw & \linethrough & \qw & \qw & \qw & \qw & \qw & \qw & \qw & \qw & \qw & \qw & \qw & \qw & \linethrough & \qw & \qw & \qw & \octrl{-1} & \qw & \qw\\
				\lstick[]{$a_2$}
				& \qw & \linethrough & \qw & \qw & \qw & \qw & \gate{R_y} & \gate{R_y} & \gate{R_y^\dagger} & \gate{R_y^\dagger} & \qw & \qw & \qw & \qw & \linethrough & \qw & \qw & \qw & \octrl{-1} & \qw & \qw\\
				\lstick[wires=3]{$\ket{0}$}
				& \qw & \qw & \qw & \qw & \qw & \qw & \qw & \qw & \qw & \qw & \qw & \qw & \qw & \qw & \qw & \qw & \qw & \qw & \qw & \qw & \qw\\
				& \vdots &&&&&&&&&&&&&&&&&&& \vdots &\\
				& \qw & \qw & \qw & \qw & \qw & \qw & \qw & \qw & \qw & \qw & \qw & \qw & \qw & \qw & \qw & \qw & \qw & \qw & \qw & \qw & \qw
			\end{quantikz}
		};
	\end{tikzpicture}
\end{equation*}

\newpage

\noindent
Therefore, a naive counting of Toffoli gates consumed by a controlled-$U_H^\ast$ is a sum of
\begin{itemize}
	\item two $O_F$'s: $2 \times 4n$
	\item two multi-controlled-$O_H^{\text{mod.}}$'s: $2 \times \Big[2\times 2n + 2 \times 2\big\{(n+3) - 1\big\}\Big]$
	\item two multi-controlled-``swap''s: $2 \times \Big[n \times 2(3-1) \Big]$
	\item two adders: $2 \times 2n$
	\item multi-controlled $R_y$'s (suitably permuted and paired up): $2 \times 2[(n+3)-1]$
	\item a multi-controlled gate at the end: $2[(n+4)-1]$
\end{itemize}
which is equal to $42n+30$.
The results are summarized as follows:

\begin{center}
	\begin{table}[h]
		\centering
		\begin{tabular}{c|wc{50mm}wc{50mm}}
			\hline
			& \#(logical qubits)/$U_H^\ast$ & \#(Toffoli gates)/controlled-$U_H^\ast$\\
			\hline
			Example 1 & $4n+2$ & $11n+2$\\
			Example 2 & $4n+6$ & $42n+30$\\
			\hline
		\end{tabular}
		\caption{Summary of the costs necessary to solve the problems for $2^n$ point masses.}
	\end{table}
\end{center}

Note that the circuit implementations in this section are presented in a redundant form
which (is easier to understand and) extends to more complicated problems at similar cost.
As a result, they are far from optimal for the specific examples,
and there is in fact plenty of room for \emph{ad hoc} cost reduction by directly exploiting the structures of each problem.
Although it is not our primary focus to reduce the costs as much as possible for the toy models,
we present an example of fine-tuned implementation for Example 2 in Appendix~\ref{appendix:speedrun},
which consumes far fewer qubits and Toffoli gates compared to the naive implementation.

\newpage

\subsection{Generalization: multiple layers}
In passing, we also mention a more abstract level of oracle utilization.
Consider a generic system of $K$ point masses (not necessarily sparsely) connected to each other.
Then, one can fabricate oracles $O_F^{\text{full}}, O_H^{\text{full}}$ for a system of $N=K\times L$ point masses
comprised of $L$ copies of the original system layered on top of each other,
if oracles $O_F^{\text{sub}}, O_H^{\text{sub}}$ slightly modified as follows for the original system are available somehow:
For a $x$-th point mass ($x\in\{1,...,K\}$) of a $l$-th layer ($l\in\{1,...,L\}$),
\begin{equation*}
    O_F^{\text{sub}}\ket{x,i}\ket{0}=\ket{x,y_i}\ket{\Delta l_i}
\end{equation*}
where $i$ labels the spring connecting the $x$-th point mass to a $y_i$-th point mass of a $(l+\Delta l_i)$-th layer, and 
\begin{equation*}
    O_H^{\text{sub}}\ket{x,y}\ket{\Delta l}\ket{z}=\ket{x,y}\ket{\Delta l}\ket{z\oplus H_{\Delta l, xy}}
\end{equation*}
where $H_{\Delta l\neq 0,xy}$ describes connections between different layers.
For example for the simplest case, $\Delta l_i$ is either $-1,0,$ or $+1$,
and the underlying matrices satisfy $H_{-1,xy} = -H_{+1,yx}$.

The target oracle $O_F^{\text{full}}$ for the full system
\begin{equation*}
	O_F^{\text{full}}\ket{x,i}\ket{l,0}=\ket{x,y_i}\ket{l,l+\Delta l_i}
\end{equation*}
can be realized by first applying $O_F^{\text{sub}}$ and then applying an adder.
Similarly, the oracle $O_H^{\text{full}}$ for the full system
\begin{equation*}
	O_H^{\text{full}} \ket{x,y}\ket{l,l+\Delta l}\ket{z}
	=
	\ket{x,y}\ket{l,l+\Delta l}\ket{z\oplus H_{\Delta l, xy}}
\end{equation*}
can be realized just by applying $O_H^{\text{sub}}$.
This allows us to analyze large systems with repetitive structure,
which we expect to be in high demand for many practical applications.
In particular, assuming that each layer is connected to a constant (i.e. $L$-independent) number of other layers
(so that $H_{\Delta}\neq 0$ only for a limited number of $\Delta l$'s, which is often the case),
this construction is exponentially efficient in terms of $L$, the number of repetitions,
as dependence on it arises only from the adder of $O_F^{\text{full}}$.
Also, note that the previous examples can be regarded as special instances of this generic construction
where $(K,L) = (1,2^n)$ or $(2, 2^{n-1})$ (instead of $(2^n, 1)$; original $\ket{x,y}$ corresponds to $\ket{l,l+\Delta l}$ from this point of view).

\newpage

\section{Initial states and their preparation}\label{sec:state_preparation}
The remaining necessary ingredient to carry out a quantum phase estimation algorithm is
an initial state upon which a sequence of powers-of-unitary is applied.
If one can prepare an exact eigenstate $\ket{\psi_{\text{exact}}}$ of the unitary, 
then the phase estimation outputs the desired eigenvalue
after a single iteration of the whole circuit.
On the other hand, if one can only prepare an approximate eigenstate $\ket{\psi_{\text{approx.}}}$ of the unitary instead,
then the success probability of QPE is given by the overlap $|\langle \psi_{\text{approx.}}|\psi_{\text{exact}}\rangle|^2$,
meaning that the expectation value of the number of iterations necessary to extract a correct eigenvalue is its reciprocal.
While it is almost impossible to prepare the ideal state $\ket{\psi_{\text{exact}}}$ in general,
it is often assumed that some \textit{nice} approximate state $\ket{\psi_{\text{approx.}}}$ can be efficiently prepared somehow,
without ruining the exponential speedup over classical computation.
In this section, 
we propose a naive way of state preparation for the problems of interest,
and examine its effectiveness.

\subsection{Proposal}
Here we adopt an approximate eigenvector of the $2^n\times 2^n$ matrix $H$
as the input state to a QPE algorithm, ``approximating'' the eigenstate \eqref{eq:eigenvector_W} of the unitary $U_H^\ast$.\footnote{
	So the success probability of the QPE algorithm is about $0.5$ (which is large enough for our purpose).
}
One way to prepare it is the following:
take small enough $2^m$ so that one can classically compute (within reasonable time) the exact eigenvector $\ket{\psi_m}$ of an ``approximate $H$'' whose size is $2^m\times 2^m$.
Encoding this state to $m$ qubits by hand
and then adding $(n-m)$ qubits set to $\ket{+}$,
one can create an $n$-qubit state
\begin{equation*}
	\ket{\psi_n^{(m)}}
	\coloneqq
	\ket{\psi_m}
	\otimes
	\dfrac{\ket{00\dots0}
		+\dots
		+\ket{11\dots 1}
	}{
		\sqrt{2}^{n-m}
	}
\end{equation*}
which serves as an approximation to the eigenvector $\ket{\psi_n}$ of the target $2^n\times 2^n$ matrix $H$.
For example, for $m=2$ and $n=4$, the procedure can be expressed as
\begin{equation*}
	\begin{array}{ccrc}
		\ket{\psi_m} = a\ket{\bm{00}} + b\ket{\bm{01}} + c\ket{\bm{10}} + d\ket{\bm{11}}
		& \mapsto & \ket{\psi_n^{(m)}} = & \frac{a}{2}(\ket{\bm{00}00} + \ket{\bm{00}01} + \ket{\bm{00}10} + \ket{\bm{00}11})\\
		&& + & \frac{b}{2}(\ket{\bm{01}00} + \ket{\bm{01}01} + \ket{\bm{01}10} + \ket{\bm{01}11})\\
		&& + & \frac{c}{2}(\ket{\bm{10}00} + \ket{\bm{10}01} + \ket{\bm{10}10} + \ket{\bm{10}11})\\
		&& + & \frac{d}{2}(\ket{\bm{11}00} + \ket{\bm{11}01} + \ket{\bm{11}10} + \ket{\bm{11}11}).\\
	\end{array}
\end{equation*}
The point of this method is that
the desired approximate state can be prepared in an exponentially efficient manner in the sense that
it only involves insertions of exponentially small number of qubits compared to the size (i.e. $2^n$) of the state.

\newpage

\subsection{Numerical experiments}
To check how well this method approximates the exact state,
we compute and compare approximate eigenvectors $\ket{\psi_n^{(m)}}$ for relatively small $2^n$.
As a naive choice, here we just take the $2^m\times 2^m$-matrix version of Eq.\,\eqref{eq:2-sparse-matrix} as an ``approximate $H$'';
for Example 1, the results are as follows:
\begin{figure}[h!]
	\centering
	\includegraphics[width=0.54\textwidth]{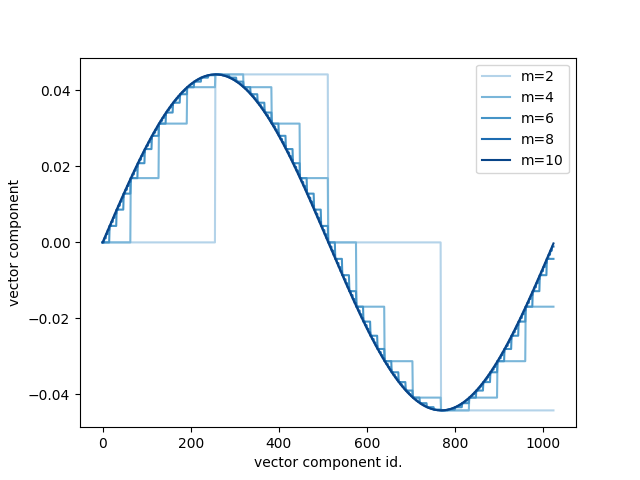}
	\caption{
		Approximate fundamental eigenvectors $\ket{\psi_n^{(m)}}$ of
		the matrices \eqref{eq:Hamiltonian_ex1} of Example~1 for $n=10$, rescaled and aligned.
	}
\end{figure}
\begin{figure}[h!]
	\centering
	\includegraphics[width=0.54\textwidth]{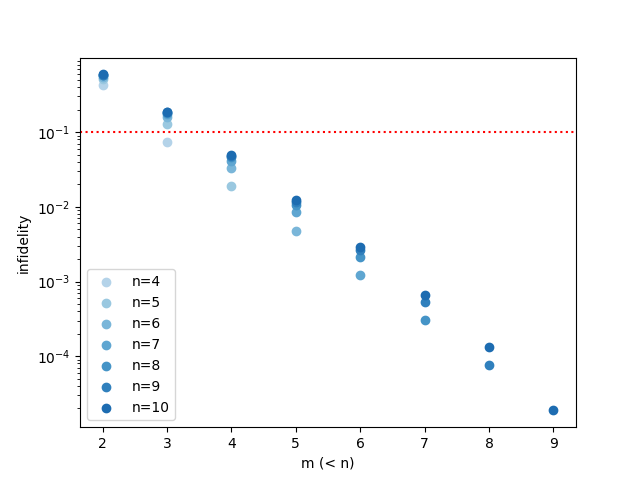}
	\caption{
		Infidelity $1 - |\langle \psi_{n} | \psi_{n}^{(m)}\rangle|^2$ between
		the exact fundamental eigenvectors $\ket{\psi_{n}}$ and approximate eigenvectors $|\psi_{n}^{(m)}\rangle$
		of the matrices \eqref{eq:Hamiltonian_ex1}.
		The red dotted horizontal line denotes infidelity $0.1$, implying that $m=3$ is already a \emph{good} approximation.
	}
\end{figure}

\newpage

\noindent
For Example 2, the results are basically the same, 
but one might want to consider the matrices \eqref{eq:Hamiltonian_ex1} with $k=\frac{k_1+k_2}{2}$ instead
in order to regard the approximation as \emph{coarse graining}.

In either case, eigenvectors for very small $2^m$ already seem to approximate those for large $2^n$ quite well.
In fact, the infidelity $1 - |\langle \psi_{n} | \psi_{n}^{(m)}\rangle|^2$ is about $0.1$ for $m=3$ for any $n \,(\leq 10)$,
which is reasonably small and the corresponding \emph{leaven} state $\ket{\psi_m}$ is easy to prepare.
One naive way to do this is by $(2^m-1)$ sequential (controlled) rotations as
\begin{equation*}
	\begin{tikzpicture}
		\node[scale=1] at (0,0) {
			\begin{quantikz}[row sep={20pt,between origins}, column sep=10pt]
				\lstick{$\ket{0}$} & \qw & \gate{R_y(\theta_0)} & \octrl{1} & \ctrl{1} & \octrl{1} & \octrl{1} & \ctrl{1} & \ctrl{1} & \qw\ \cdots\ & \qw & \qw & \qw\\
				\lstick{$\ket{0}$} & \qw & \qw & \gate{R_y(\theta_1)} & \gate{R_y(\theta_2)} & \octrl{1} & \ctrl{1} & \octrl{1} & \ctrl{1} & \qw\ \cdots\ & \qw & \qw & \qw\\
				\lstick{$\ket{0}$} & \qw & \qw & \qw & \qw & \gate{R_y(\theta_3)} & \gate{R_y(\theta_4)} & \gate{R_y(\theta_5)} & \gate{R_y(\theta_6)} & \qw\ \cdots\ & \qw & \qw & \qw\\
				& \vdots &&&&&&&& \ddots\\
			\end{quantikz}
		};
	\end{tikzpicture}
\end{equation*}
where for $\ket{\psi_m} = \displaystyle \sum_{i=0}^{2^m-1}a_i \ket{i}$ $(a_i \in \mathbb{R})$
the angles $\{\theta_i\}$ are precomputed as
\begin{equation*}
	\begin{array}{lcl}
		\theta_0 & = & 2\arccos\sqrt{\tfrac{a_0^2+a_1^2+\cdots + a_{2^{m-1}-1}^2}{a_0^2+a_1^2+\cdots + a_{2^{m\phantom{-1}}-1}^2}},\\[10pt]
		\theta_1 & = & 2\arccos\sqrt{\tfrac{a_0^2+a_1^2+\cdots + a_{2^{m-2}-1}^2}{a_0^2+a_1^2+\cdots + a_{2^{m-1}-1}^2}},\\[10pt]
		\theta_2 & = & 2\arccos\sqrt{\tfrac{a_{2^{m-1}}^2+a_{2^{m-1}+1}^2+\cdots + a_{2^{m-1}+2^{m-2}-1}^2}{a_{2^{m-1}}^2+a_{2^{m-1}+1}^2+\cdots + a_{2^{m\phantom{-1}}\phantom{+2^{m-2}}-1}^2}},
	\end{array}
\end{equation*}
and so on,
but one can also use more intricate circuits such as the PREPARE \cite{BabbushGidneyBerryWiebeMcCleanPalerFowlerNeven:2018},
at the expense of additional qubits and gates.

\newpage

\section{Resource estimation}\label{sec:resource_estimation}

In order to implement a quantum phase estimation algorithm on actual quantum devices,
one has to employ quantum error-correcting codes and make the circuit fault-tolerant.
One of the popular schemes used in previous resource estimation literature is
the \emph{surface code} \cite{Kitaev:1997,BravyiKitaev:1998,DennisKitaevLandahlPreskill:2001}
based on the \emph{lattice surgery} \cite{HorsmanFowlerDevittVanMeter:2011},
as it achieves relatively high threshold error rates
using only local interactions between nearest-neighbor qubits on an array,
allowing it to possess high modularity.

The overhead due to this process is mainly twofold:
encoding of logical qubits of the quantum circuit into physical qubits,
and distillation of  the \emph{magic states} to be consumed to realize non-Clifford gates \cite{BravyiKitaev:2004}.
In this section, we examine them in order and give a very rough estimation\footnote{
	Precise estimation involves innumerable subtleties (e.g. execution of Clifford gates, routing of qubits), 
	and here we turn a blind eye to them and content ourselves with a superficial analysis.
	The results are hopefully within an order of magnitude of the true values.
} of the number of necessary physical qubits and actual runtime,
following \cite{Litinski:2018,Litinski:2019}.
Notations and details of parameters are summarized below.

\begin{center}
	\begin{tabular}{wc{30mm}lwc{50mm}}
		\hline
		parameter & \multicolumn{1}{c}{description} & \multicolumn{1}{c}{assumed value}\\
		\hline
		$\epsilon_{\text{prec.}}$ & required precision of eigenvalues & $10^{-4}$\\
		$\epsilon_{\text{fail}}$ & target failure rate of the whole circuit & $10^{-2}$\\
		$p_{\text{phys.}}$ & physical error rate of elementary operations & $10^{-3}$\\
		$t_{\text{cycle}}$ & code cycle time & $1$ $\mu$s\\
		\hline
	\end{tabular}
\end{center}

\subsection{Encoding}

Given an $n_{\text{logical}}$-qubit quantum circuit, 
one first needs to store the qubits in a larger \emph{block} designed to enable them to consume magic states via lattice surgery.
There is a trade-off between the block size $b(n_{\text{logical}})$ and the necessary code cycles $c_{\text{consume}}$ for a magic state consumption,
and the two extreme cases taken up in \cite{Litinski:2018} are the following:
\begin{center}
		\renewcommand{\arraystretch}{1.44}
		\begin{tabular}{c|cc}
			\hline
			block type & $b(n_{\text{logical}})$ & $c_{\text{consume}}$\\
			\hline
			compact & $\big\lceil \frac{n_{\text{logical}} + 2}{2} \big\rceil \times 3$ & $9d_{\text{code}}$\\
			fast & $\displaystyle\min_{1\leq k \leq n_{\text{logical}}} (2k+1)\cdot (\lceil\tfrac{n_{\text{logical}}}{k}\rceil+1)$ & $d_{\text{code}}$\\
			\hline
		\end{tabular}
\end{center}
where $d_{\text{code}}$ is the code distance.
Then, each of the $b(n_{\text{logical}})$ logical qubits is converted to $d_{\text{code}}^2$ physical qubits,
equipped with another $d_{\text{code}}^2$ physical qubits used for syndrome measurements.
Since the logical error rate per logical qubit per code cycle (round) is known \cite[Fig.\,8]{FowlerWangHollenberg:2010} to be approximated as
\begin{equation*}
	p_{\text{logical}} = 0.1 \cdot (100\, p_{\text{phys.}})^{\lceil\frac{d_{\text{code}}}{2}\rceil},
\end{equation*}
the code distance $d_{\text{code}}$ to be adopted is determined by requiring
\begin{equation}\label{eq:code-distance-requirement}
	p_{\text{logical}} \cdot b(n_{\text{logical}}) \cdot \max\big(
		c_{\text{consume}},
		c_{\text{produce}}
	\big) \cdot n_{T} < \epsilon_{\text{fail}}
\end{equation}
where $c_{\text{produce}}$ is the (net) necessary code cycles for production of a (distilled) magic state at \textit{factories} built out of $n_{\text{distill}}$ physical qubits in total,
and $n_T$ is the number of $T$ gates in the quantum circuit (after suitably decomposing all the non-Clifford gates).
Making a choice of $d_{\text{code}}$ satisfying Eq.\,\eqref{eq:code-distance-requirement},
one can calculate the total number of necessary physical qubits and actual runtime as
\begin{equation*}
	\begin{array}{lcl}
		n_{\text{phys.}} & = & b(n_{\text{logical}}) \cdot 2d_{\text{code}}^2 + n_{\text{distill}},\\[4pt]
		t_{\text{total}} & = & \max\big(
			c_{\text{consume}},
			c_{\text{produce}}
		\big) \cdot n_{T} \cdot t_{\text{cycle}}.\\[10pt]
	\end{array}
\end{equation*}
For the analysis below however,
we will focus on Toffoli gates instead of $T$ gates and correspondingly the factor $\max\big(
	c_{\text{consume}},
	c_{\text{produce}}
\big) \cdot n_{T}$ is suitably replaced.

\subsection{Realization of Toffoli gates}
One way to execute a Toffoli gate is by consuming a Toffoli or CCZ state
which can be synthesized from magic states \cite{Eastin:2012,Jones:2012,CampbellHoward:2016-2,CampbellHoward:2016-1}.
According to \cite{Litinski:2019},
one can construct a \emph{factory} consisting of $0.5\times 10^5$ physical qubits,
which can ship a CCZ state with small enough infidelity ($< 10^{-10}$) every $60$ code cycles.
As one typically has $d_{\text{code}}\geq 7$ (and thus $9d_{\text{code}} > 60$),
we adopt the \emph{fast block} above for encoding and operate $n_{\text{factory}}$ factories so that
\begin{equation*}
	\begin{array}{lcl}
		n_{\text{distill}} & = & 0.5\times 10^5 \cdot n_{\text{factory}},\\[4pt]
		\max\big(
			c_{\text{consume}},
			c_{\text{produce}}
		\big)
		& = & 
		\max\big(
			d_{\text{code}}, \frac{60}{n_{\text{factory}}}
		\big).
	\end{array}
\end{equation*}

\newpage

\subsection{Case study}
Let us take up the naive implementations of examples in Sec.\,\ref{sec:main_analysis}, expecting that
required costs are not that different from those of slightly more general problems and thus give indications thereof.
The quantum circuit to start with is that of quantum phase estimation with unitary $U_H^\ast$ described in Sec.\,\ref{subsec:ex1} or Sec.\,\ref{subsec:ex2} plugged in,
where for the latter one has
\begin{equation*}
	\renewcommand{\arraystretch}{1.5}
	\begin{array}{lcl}
		n_{\text{logical}} & = & (4n+6) + \left\lceil\log_2\tfrac{1}{\epsilon_{\text{prec.}}}\right\rceil,\\
		n_{\text{Toffoli}} & = & (42n+30) \times (2^{\lceil\log_2\frac{1}{\epsilon_{\text{prec.}}}\rceil} - 1).
	\end{array}
\end{equation*}
If one takes $\epsilon_{\text{prec.}} = 10^{-4}$ (and ignores the normalization of $H$ for simplicity), it turns out that $n_{\text{Toffoli}} \lesssim 10^8$ for $n \lesssim 128$,
and thus the CCZ state above with infidelity $< 10^{-10}$ is reasonable to use.\footnote{
	Here we also ignore the costs required for (not only Clifford gates but also)
	Pauli rotation gates in both the unitary $U_H^\ast$ and the inverse Quantum Fourier Transform component,
	whose numbers are independent of the matrix order $2^n$.
	They can be fault-tolerantly implemented by decomposition into Clifford+$T$ gates, combining the methods of \cite{RossSelinger:2014} and \cite{KliuchnikovMaslovMosca:2012},
	which roughly uses $3\log_2 \frac{1}{\epsilon_{\text{prec.}}} \cdot [4 \cdot \frac{1}{\epsilon_{\text{prec.}}} + \frac{1}{2}(\log_2 \frac{1}{\epsilon_{\text{prec.}}})^2]$ $\sim 10^6$ $T$ gates
	and should be negligible compared to the Toffoli gates' cost for large $2^n$.
}
Following the preceding argument, some rough estimates are given as follows: 
\begin{table}[h]
	\centering
	\begin{tabular}{c|wc{10mm}wc{50mm}wc{35mm}wc{20mm}}
		\hline
		order of matrix & $d_{\text{code}}$ & \# of logical/physical qubits & \# of Toffoli gates & runtime\\
		\hline
		$2^{32\phantom{0}}$ & $23$ & $\phantom{0}330$ / $\sim \phantom{1}4.5\times 10^5$ & $\sim 0.6 \times 10^7$ & $\sim 6$ min.\\
		$2^{64\phantom{0}}$ & $25$ & $\phantom{0}594$ / $\sim \phantom{1}8.4\times 10^5$ & $\sim 1.2 \times 10^7$ & $\sim 12$ min.\\
		$2^{128}$ & $25$ & $1128$ / $\sim 15.1\times 10^5$ & $\sim 2.3 \times 10^7$ & $\sim 24$ min.\\
		\hline
	\end{tabular}
	\begin{tabular}{c|wc{10mm}wc{50mm}wc{35mm}wc{20mm}}
		\hline
		order of matrix & $d_{\text{code}}$ & \# of logical/physical qubits & \# of Toffoli gates & runtime\\
		\hline
		$2^{32\phantom{0}}$ & $25$ & $\phantom{0}336$ / $\sim \phantom{1}5.2\times 10^5$ & $\sim 2.3 \times 10^7$ & $\sim 23$ min.\\
		$2^{64\phantom{0}}$ & $25$ & $\phantom{0}608$ / $\sim \phantom{1}8.6\times 10^5$ & $\sim 4.5 \times 10^7$ & $\sim 45$ min.\\
		$2^{128}$ & $27$ & $1134$ / $\sim 17.5\times 10^5$ & $\sim 8.9 \times 10^7$ & $\sim 90$ min.\\
		\hline
	\end{tabular}
	\caption{
		Summary of very rough estimates of the costs necessary to solve the problems of	Example 1 (above) and 2 (below) for $2^n$ point masses,
		$\epsilon_{\text{prec.}}=10^{-4}$, $t_{\text{cycle}}=1\,\mu$s (and $k_2 = \frac{1}{2}$ for the latter),
		using the fast block of \cite{Litinski:2018} and $n_{\text{factory}}=2$ factories of \cite{Litinski:2019} mentioned in the previous subsections.
		To estimate the runtime, we also assumed that the QPE is run twice (since the success probability was about 0.5 as mentioned in Sec.\,\ref{sec:state_preparation}).
	}
\end{table}

\noindent
We again stress that they are given only as a demonstration
with many finer points swept under the rug,
and the precise values themselves have no actual meaning at all.\footnote{
	Also, if one \emph{really} wants to do the computation for this specific problem,
	it would be better to use a more efficient circuit, for example the one described in Appendix \ref{appendix:speedrun},
	while the cost of the rotation gates might not be negligible even for $n = 128$.
}
The interested reader is referred to e.g.\,\cite{GidneyFowler:2019,LeeBerryGidneyHugginsMcCleanWiebeBabbush:2020} and references therein
for more refined analyses using CCZ state factories.


\section*{Acknowledgments}
The authors would like to thank Yuya O. Nakagawa and Shoichiro Tsutsui for careful reading of and comments on the draft.
The authors are also deeply grateful to Christoph S\"underhauf for kindly pointing out that
the method of \cite{SunderhaufCampbellCamps:2023} can be applied to our toy problems and sharing an explicit circuit construction
with far fewer qubits and Toffoli gates (and even smaller subnormalization) based on it,
which inspired a further-improved implementation by the authors presented in Appendix~\ref{appendix:speedrun}.

\newpage

\appendix
\renewcommand{\thesubsection}{\thesection.\arabic{subsection}}
\section{Quantum arithmetic and related operations}\label{appendix:arithmetic}

We summarize the necessary modules
to decompose a full circuit into elementary gates.

\subsection{Adder}\label{subsec:Adder}
It is known that a sum (modulo $2^n$) of two $n$-bit integers encoded in quantum states can be computed
using only a single ancillary qubit \cite{CuccaroDraperKutinMoulton:2004}.
The basic strategy is to first compute all the \textit{carries}, and then do the summation.
Correspondingly, the adder consists of two elementary components.

The first component is a \emph{majority} gate, implemented as
\begin{equation*}
	\begin{tikzpicture}
		\node[scale=1.1] at (0,0) {
			\begin{quantikz}[row sep={20pt,between origins}, column sep=10pt]
				\lstick[]{$\ket{c}$} & \gate[wires=3]{\text{majority}} & \qw \rstick[]{$\ket{\ast}$}\\
				\lstick[]{$\ket{a}$} & \qw & \qw \rstick[]{$\ket{\ast}$}\\
				\lstick[]{$\ket{b}$} & \qw & \qw \rstick[]{$\ket{\texttt{majority}(a,b,c)}$}
			\end{quantikz}
		};
		\node[] at (4.5,0) {$=$};
		\node[scale=1.1] at (8.8,0) {
			\begin{quantikz}[row sep={20pt,between origins}, column sep=10pt]
				\qw & \qw		& \targ{}   & \ctrl{1} & \qw \rstick[]{$\ket{c \oplus b}$}\\
				\qw & \targ{}	& \qw		& \ctrl{1} & \qw \rstick[]{$\ket{a \oplus b}$}\\
				\qw & \ctrl{-1} & \ctrl{-2} & \targ{} & \qw \rstick[]{$\ket{\texttt{majority}(a,b,c)}$}
			\end{quantikz}
		};
	\end{tikzpicture}
\end{equation*}
where $a,b \in \{0,1\}$ correspond to summand bits and $c \in \{0,1\}$ corresponds to a carry from lower bits.
The purpose of the gate is to output a carry of the addition $c+a+b$, which is equal to $\texttt{majority}(a,b,c)$.

If $b=0$, one does not have to do anything to the corresponding qubit $\ket{b}$,
except when $\texttt{majority}(a,b,c)=1$, i.e. when both $c$ and $a$ are $1$,
which can be realized by a flipping of $b$ conditioned on $c=a=1$.
This explains the rightmost Toffoli gate.
On the other hand, if $b=1$, one has to flip this bit when $\texttt{majority}(a,b,c)=0$ i.e. when both $c$ and $a$ are $0$,
and one can utilize the same Toffoli gate with two preceding CNOT gates to achieve this.

The second component is an \emph{un-majority + add} gate, implemented as
\begin{equation*}
	\begin{tikzpicture}
		\node[scale=1.1] at (0,0) {
			\begin{quantikz}[row sep={20pt,between origins}, column sep=10pt]
				\lstick[]{$\ket{c\oplus b}$} & \gate[wires=3]{\text{majority}^\dagger} &\gate[wires=3]{\text{add}} & \qw \rstick[]{$\ket{c}$}\\
				\lstick[]{$\ket{a\oplus b}$} & \qw & \qw & \qw \rstick[]{$\ket{a\oplus b\oplus c}$}\\
				\lstick[]{$\ket{\texttt{majority}(a,b,c)}$} & \qw & \qw & \qw \rstick[]{$\ket{b}$}
			\end{quantikz}
		};
		\node[] at (5.8,0) {$=$};
		\fill[black!20, opacity=0.5] (8.0,-1.1) -- (9.6,-1.1) -- (9.6,0.4) -- (8.0,0.4) -- (8.0,-1.1);
		\node[scale=1.1] at (8.4,0) {
			\begin{quantikz}[row sep={20pt,between origins}, column sep=10pt]
				\qw & \ctrl{1} & \targ{}   & \qw 	   & \qw	   & \ctrl{1} & \qw\\
				\qw & \ctrl{1} & \qw		& \targ{}  & \targ{}   & \targ{} & \qw\\
				\qw & \targ{} & \ctrl{-2} & \ctrl{-1} & \ctrl{-1} & \qw		& \qw
			\end{quantikz}
		};
	\end{tikzpicture}
\end{equation*}
which (undoes the \emph{majority} operation and then) outputs the bitwise summation result $a\oplus b\oplus c$.
Here, the two CNOT gates in the shaded box cancel out.

\newpage

With these two components,
the desired adder of two $n$-bit integers $x,y$ is realized using only a single ancillary qubit $A$ as
\begin{equation*}
	\begin{tikzpicture}
		\node[scale=1.0] at (0,0) {
			\begin{quantikz}[row sep={20pt,between origins}, column sep=10pt]
				\lstick{$\ket{0}_{\!A}$} 		&
				\gate[wires=3]{\rotatebox[]{90}{\hspace{-20pt}$\mathrm{majority}$\hspace{-20pt}}} & \qw & \qw\ \cdots\ & \qw & \qw & \qw & \qw\ \cdots\ & \qw & \qw& \qw &
				\gate[wires=3]{\rotatebox[]{90}{\hspace{-20pt}$\mathrm{majority}^\dagger$\hspace{-20pt}}} &
				\gate[wires=3]{\rotatebox[]{90}{\hspace{-5pt}$\mathrm{add}$\hspace{-5pt}}} & \qw \rstick{$\ket{0}_{\!A}$}\\
				\lstick{$\ket{x_0}$} 		& \qw 												& \qw & \qw\ \cdots\ & \qw & \qw & \qw & \qw\ \cdots\  & \qw & \qw & \qw & \qw & \qw & \qw\rstick{$\ket{(x+y)_0}$}\\
				\lstick{$\ket{y_0}$} 		& \qw 												&
				\gate[wires=3]{\rotatebox[]{90}{\hspace{-20pt}$\mathrm{majority}$\hspace{-20pt}}} & \qw\ \cdots\  & \qw & \qw & \qw & \qw\ \cdots\  &
				\gate[wires=3]{\rotatebox[]{90}{\hspace{-20pt}$\mathrm{majority}^\dagger$\hspace{-20pt}}} &
				\gate[wires=3]{\rotatebox[]{90}{\hspace{-5pt}$\mathrm{add}$\hspace{-5pt}}} & \qw & \qw & \qw & \qw\rstick{$\ket{y_0}$}\\
				\lstick{$\ket{x_1}$} 		& \qw 												& \qw & \qw\ \cdots\ & \qw & \qw & \qw & \qw\ \cdots\  & \qw & \qw & \qw & \qw & \qw & \qw\rstick{$\ket{(x+y)_1}$}\\
				\lstick{$\ket{y_1}$} 		& \qw 												& \qw & \qw\ \cdots\ & \qw & \qw & \qw & \qw\ \cdots\ & \qw & \qw & \qw & \qw & \qw & \qw\rstick{$\ket{y_1}$}\\[10pt]
				& \vdots && \rotatebox[]{-10}{$\ddots$} &&&& \rotatebox[]{80}{$\ddots$}\\[10pt]
				\lstick{$\ket{y_{n-2}}$} & \qw 												& \qw & \qw\ \cdots\ &
				\gate[wires=3]{\rotatebox[]{90}{\hspace{-20pt}$\mathrm{majority}^{\phantom{\dagger}}$\hspace{-20pt}}} &
				\gate[wires=3]{\rotatebox[]{90}{\hspace{-20pt}$\mathrm{majority}^\dagger$\hspace{-20pt}}} &
				\gate[wires=3]{\rotatebox[]{90}{\hspace{-5pt}$\mathrm{add}$\hspace{-5pt}}} & \qw\ \cdots\ & \qw & \qw & \qw & \qw & \qw & \qw\rstick{$\ket{y_{n-2}}$}\\
				\lstick{$\ket{x_{n-1}}$} & \qw 												& \qw & \qw\ \cdots\ & & & & \qw\ \cdots\ & \qw & \qw & \qw & \qw & \qw & \qw\rstick{$\ket{(x+y)_{n-1}}$}\\
				\lstick{$\ket{y_{n-1}}$} & \qw 												& \qw & \qw\ \cdots\ & & & & \qw\ \cdots\ & \qw & \qw & \qw & \qw & \qw & \qw\rstick{$\ket{y_{n-1}}$}
			\end{quantikz}
		};
	\end{tikzpicture}
\end{equation*}
where the top qubit of each majority gate always encodes a carry and the other two qubits encode summands.
Here, a naive counting tells us that the number of necessary Toffoli gates for each full adder of $n$-bit integers is $2n$.\footnote{
	The number of gates can be \emph{halved} (although the count is in terms of $T$ gates) at the cost of additional $n$ ancillary qubits (and their measurements) \cite{Gidney:2018},
	but for the sake of simplicity we will stick to the naive adder explained above.
}

Also, note that a controlled adder can be realized using $4n$ Toffoli gates,
thanks to the trick mentioned in Appendix \ref{subsec:Controlled-U}; one just needs to control
the first CNOT gates of majority gates and
the last CNOT gates of un-majority\,+\,add gates,
as the remaining gates are in pairs and need not be controlled.

\newpage

\subsection{Controlled unitaries}\label{subsec:Controlled-U}
When adding control qubits to a large quantum gate which consists of multiple (smaller) gates,
there is a useful trick to keep in mind;
if there exists a subsequence of gates which as a whole constitutes a trivial gate,
one does not have to add control qubits to those gates
as the overall action does not change.

An especially important example is the following:
As is well-known (see e.g. a standard textbook \cite{NielsenChuang:2000}), using three unitaries
\begin{equation*}
	\begin{array}{ccl}
		A & = & R_z(\beta) \,R_y(\frac{\gamma}{2}),\\[3pt]
		B & = & R_y(-\frac{\gamma}{2})\,R_z(-\frac{\beta+\delta}{2}),\\[3pt]
		C & = & R_z(-\frac{\beta-\delta}{2}),\\
	\end{array}
\end{equation*}
one can construct an arbitrary single-qubit unitary (up to phase) as
\begin{equation*}
	U=AXBXC=
	\left(
		\begin{array}{rr}
			e^{-i\frac{\beta+\delta}{2}}\cos \frac{\gamma}{2} &
			-e^{-i\frac{\beta-\delta}{2}}\sin \frac{\gamma}{2}\\[3pt]
			e^{+i\frac{\beta-\delta}{2}}\sin \frac{\gamma}{2} &
			e^{+i\frac{\beta+\delta}{2}}\cos \frac{\gamma}{2}
		\end{array}
	\right).
\end{equation*}
Given this form of decomposition, a controlled-$U$ gate can be constructed as
\begin{equation*}
	\begin{tikzpicture}
			\node[scale=1.1] at (0,0) {
				\begin{quantikz}[row sep={30pt,between origins}, column sep=10pt]
					\qw & \ctrl{1} & \qw\\
					\qw & \gate{U} & \qw
				\end{quantikz}
			};
			\node[] at (1.4,0) {$=$};
			\node[scale=1.1] at (4.5,0) {
				\begin{quantikz}[row sep={30pt,between origins}, column sep=10pt]
					\qw & \qw & \ctrl{1} & \qw & \ctrl{1} & \qw & \qw\\
					\qw & \gate{C} & \targ{} & \gate{B} & \targ{} & \gate{A} & \qw
				\end{quantikz}
			};
	\end{tikzpicture}
\end{equation*}
by adding control qubits only to $X$ gates, since $ABC = I$.
Of particular interests are the following controlled-rotations:
\begin{equation*}
	\begin{tabular}{>{$}c<{$}|>{$}wc{12mm}<{$}>{$}wc{12mm}<{$}>{$}wc{12mm}<{$}}
		U & \beta & \gamma & \delta\\
		\hline
		R_y(\theta) & 0 & \theta & 0\\
		R_z(\theta) & x & 0 & \theta-x\\
	\end{tabular}
\end{equation*}
where one can take $x = \frac{\theta}{2}$ for convenience.
\begin{equation*}
	\begin{tikzpicture}
			\node[scale=1.1] at (0,0) {
				\begin{quantikz}[row sep={30pt,between origins}, column sep=10pt]
					\qw & \ctrl{1} & \qw\\
					\qw & \gate{R_y(\theta)} & \qw
				\end{quantikz}
			};
			\node[] at (1.9,0) {$=$};
			\node[scale=1.1] at (5.5,0) {
				\begin{quantikz}[row sep={30pt,between origins}, column sep=10pt]
					\qw & \ctrl{1} & \qw & \ctrl{1} & \qw & \qw\\
					\qw & \targ{} & \gate{R_y(-\frac{\theta}{2})} & \targ{} & \gate{R_y(\frac{\theta}{2})} & \qw
				\end{quantikz}
			};
	\end{tikzpicture}
\end{equation*}
\begin{equation*}
	\begin{tikzpicture}
			\node[scale=1.1] at (0,0) {
				\begin{quantikz}[row sep={30pt,between origins}, column sep=10pt]
					\qw & \ctrl{1} & \qw\\
					\qw & \gate{R_z(\theta)} & \qw
				\end{quantikz}
			};
			\node[] at (1.9,0) {$=$};
			\node[scale=1.1] at (5.5,0) {
				\begin{quantikz}[row sep={30pt,between origins}, column sep=10pt]
					\qw & \ctrl{1} & \qw & \ctrl{1} & \qw & \qw\\
					\qw & \targ{} & \gate{R_z(-\frac{\theta}{2})} & \targ{} & \gate{R_z(\frac{\theta}{2})} & \qw
				\end{quantikz}
			};
	\end{tikzpicture}
\end{equation*}

\newpage

\subsection{Multi-controlled unitaries}\label{subsec:Multi-controlled}
A multi-controlled gate can be immediately decomposed into a single-controlled gate
sandwiched by cascaded Toffoli gates:
\begin{equation*}
	\begin{tikzpicture}
		\node[scale=1.1] at (0,0) {
			\begin{quantikz}[row sep={20pt,between origins}, column sep=10pt]
				\qw & \ctrl{1} & \qw\\
				\qw & \ctrl{1} & \qw\\
				\qw & \ctrl{1} & \qw\\
				\qw & \ctrl{1} & \qw\\
				& \vdots\\
				\qw & \ctrl{6} & \qw\\
				\\
				\\
				\\
				\\
				\\
				\qw & \gate{U} & \qw\\
			\end{quantikz}
		};
		\node at (2.5,0) {$=$};
		\node[scale=1.1] at (9,0) {
			\begin{quantikz}[row sep={20pt,between origins}, column sep=10pt]
			 & \qw & \ctrl{6} & \qw & \qw & \qw & \qw & \qw & \qw & \qw & \qw & \qw & \ctrl{6} & \qw\\
			 & \qw & \ctrl{5} & \qw & \qw & \qw & \qw & \qw & \qw & \qw & \qw & \qw & \ctrl{5} & \qw\\
			 & \qw & \qw & \ctrl{5} & \qw & \qw & \qw & \qw & \qw & \qw & \qw & \ctrl{5} & \qw & \qw \\
			 & \qw & \qw & \qw & \ctrl{5} & \qw & \qw & \qw & \qw & \qw & \ctrl{5} & \qw & \qw & \qw\\
			 & \vdots &&&& \ddots &&&& \rotatebox[]{75}{$\ddots$}\\
			 & \qw & \qw & \qw & \qw & \qw & \ctrl{5} & \qw & \ctrl{5} & \qw & \qw & \qw &  \qw & \qw\\
			 \lstick{$\ket{0}_a$} & \qw & \targ{} & \ctrl{1} & \qw & \qw & \qw & \qw & \qw & \qw & \qw & \ctrl{1} & \targ{} & \qw\\
			 \lstick{$\ket{0}_a$} & \qw & \qw & \targ{} & \ctrl{1} & \qw & \qw & \qw & \qw & \qw & \ctrl{1} & \targ{} & \qw & \qw\\
			 \lstick{$\ket{0}_a$} & \qw & \qw & \qw & \targ{} \qw & \qw & \qw & \qw & \qw & \qw & \targ{} & \qw & \qw & \qw\\
			 & \vdots &&&& \ddots & \bullet && \bullet & \rotatebox[]{75}{$\ddots$}\\
			 \lstick{$\ket{0}_a$} & \qw & \qw & \qw & \qw & \qw & \targ{} & \ctrl{1} & \targ{} & \qw & \qw & \qw & \qw\\
			 & \qw & \qw & \qw & \qw & \qw & \qw & \gate{U} & \qw & \qw & \qw & \qw & \qw\\
			\end{quantikz}
		};
	\end{tikzpicture}
\end{equation*}
In other words,
a unitary gate with $n$ control qubits can be implemented at the cost of $2(n-1)$ Toffoli gates
and $(n-1)$ ancillary qubits.

\subsection{SWAP}\label{subsec:swap}
A SWAP gate is also easy to decompose; it consists of three CNOT gates:
\begin{equation*}
	\begin{tikzpicture}
		\node[scale=1.1] at (0,0) {
			\begin{quantikz}[row sep={30pt,between origins}, column sep=10pt]
				\qw & \swap{1} & \qw\\
				\qw & \targX{} & \qw
			\end{quantikz}
		};
		\node[] at (1.4,0) {$=$};
		\node[scale=1.1] at (3.5,0) {
			\begin{quantikz}[row sep={30pt,between origins}, column sep=10pt]
				\qw & \targ{} & \ctrl{1} & \targ{} & \qw\\
				\qw & \ctrl{-1} & \targ{} & \ctrl{-1} & \qw
			\end{quantikz}
		};
	\end{tikzpicture}
\end{equation*}
The first two CNOT gates swaps the two qubits if the initial values are $(0,1)$,
while the last two CNOT gates swaps the two qubits if the initial values are $(1,0)$.
Using the trick mentioned in Appendix \ref{subsec:Controlled-U},
a (multi-)controlled SWAP gate is realized by adding corresponding control qubits only to the middle CNOT gate.

\newpage

\section{Speedrun}\label{appendix:speedrun}
Here we give a fine-tuned circuit implementation of Example 2,
which requires far fewer qubits and Toffoli gates compared to the naive one in Sec.\,\ref{subsec:ex2}.
It is inspired by a construction \`a la \cite{SunderhaufCampbellCamps:2023} provided by Christoph S\"underhauf,
and essentially relies on a direct use of the index indicating the non-zero elements,
without explicitly turning to the full row/column indices by the oracle $O_F$ of Sec.\,\ref{subsec:O_F}.

If one can construct a unitary $U_2$ such that
\begin{equation*}
	\small
	U_2\ket{x}_{x}\ket{0}_{a_1}\ket{0}_{a_2}
	=
	\left\{
		\begin{array}{ll}
			\dfrac{1}{\sqrt{2}}\bigg[
		 		\ket{x-1}_{s}\ket{0}_{a_1}
		 		\Big(\dfrac{k_1}{k_2}\ket{0} + \sqrt{1 - \Big(\dfrac{k_1}{k_2}\Big)^2}\ket{1}\Big)_{a_2}
		 		+
				\ket{x+1}_{s}\ket{1}_{a_1} \ket{0}_{a_2}
			\bigg]
			& (\text{$x$ even})\\[12pt]
			\dfrac{1}{\sqrt{2}}\bigg[
		 		\ket{x+1}_{s}\ket{0}_{a_1}
		 		\Big(\dfrac{k_1}{k_2}\ket{0} + \sqrt{1 - \Big(\dfrac{k_1}{k_2}\Big)^2}\ket{1}\Big)_{a_2}
		 		+
				\ket{x-1}_{s}\ket{1}_{a_1} \ket{0}_{a_2}
			\bigg]
			& (\text{$x$ odd})
		\end{array}
	\right.
\end{equation*}
then taking $U_1$ to be an Hadamard gate acting on the ancilla qubit $a_1$
\begin{equation*}
	\small
	U_1\ket{y}_{s}\ket{0}_{a_1}\ket{0}_{a_2}
	=
	\ket{y}_{s} \frac{\ket{0}_{a_1}+\ket{1}_{a_1}}{\sqrt{2}} \ket{0}_{a_2},
\end{equation*}
the product $U_2^\dagger U_1$ block-encodes the original matrix $H$ \eqref{eq:Hamiltonian_ex2} as
\begin{equation*}
	\small
	\bra{x}_{s}\bra{0}_{a_1}\bra{0}_{a_2} U_2^\dagger U_1 \ket{y}_{s}\ket{0}_{a_1}\ket{0}_{a_2}
	\ \sim\  
	\left\{
		\begin{array}{cl}
			\frac{k_1}{k_2}\ /\ 1 & (\text{``$x$ even, $x-y=+1\,/-\!1$''}),\\[4pt]
			1\ /\ \frac{k_1}{k_2} & (\text{``$x$ odd, \ $x-y=+1\,/-\!1$''}),\\[4pt]
			0 & (\text{otherwise}).
		\end{array}
	\right.
\end{equation*}
Starting from a state $\ket{x}_{s}\ket{0}_{a_1}\ket{0}_{a_2}$,
one can see that $U_2$ can be realized for example by sequentially acting
\begin{enumerate}
	\item an Hadamard gate acting on $a_1$,\\[4pt]
	making an equal-superposition state {\small $\displaystyle \frac{1}{\sqrt{2}} \Big(\ket{x}_{s}\ket{0}_{a_1}\ket{0}_{a_2} + \ket{x}_{s}\ket{1}_{a_1}\ket{0}_{a_2}\Big)$},\\[-14pt]
	\item an adder with suitable preprocessing and uncomputation,\\[4pt]
	turning the state into {\small $\displaystyle \frac{1}{\sqrt{2}} \Big(\ket{x-1}_{s}\ket{0}_{a_1}\ket{0}_{a_2} + \ket{x+1}_{s}\ket{1}_{a_1}\ket{0}_{a_2}\Big)$},\\[-14pt]
	\item a CNOT gate controlled on the parity of $x$ (i.e. the least significant (qu)bit of $s$) and targeting on $a_1$, leading to
	\begin{equation*}
		\small
		\left\{
			\begin{array}{ll}
				\dfrac{1}{\sqrt{2}}\Big(
			 		\ket{x-1}_{s}\ket{0}_{a_1}\ket{0}_{a_2}
			 		+
					\ket{x+1}_{s}\ket{1}_{a_1} \ket{0}_{a_2}
				\Big)
				& (\text{$x$ even})\\[12pt]
				\dfrac{1}{\sqrt{2}}\Big(
			 		\ket{x+1}_{s}\ket{0}_{a_1}\ket{0}_{a_2}
			 		+
					\ket{x-1}_{s}\ket{1}_{a_1} \ket{0}_{a_2}
				\Big)
				& (\text{$x$ odd})
			\end{array}
		\right.
	\end{equation*}
	\item a controlled rotation controlled on $a_1$ and targeting on $a_2$, achieving the desired state.
\end{enumerate}

\newpage

\begin{equation*}
	\begin{tikzpicture}
		\node[scale=0.9] at (0,0) {
			\begin{quantikz}[row sep={20pt,between origins}, column sep=10pt]
				\lstick[4]{$\ket{0}$} & \qw & \targ{} & \gate[wires=8, nwires={2,6}]{\ \ \text{adder}\ \ } & \targ{} & \qw & \qw & \qw & \qw\\
				& \vdots & \vdots && \vdots &&& \vdots &\\
				& \qw & \targ{} & \qw & \targ{} & \qw & \qw & \qw & \qw\\
				& \targ{} & \qw & \qw & \qw & \targ{} & \qw & \qw& \qw \\
				\lstick[4]{$\ket{x}_{s}$} & \qw & \qw & \gateoutput[4]{$x\pm 1$} & \qw & \qw & \qw & \qw & \qw\\
				& \vdots &&&&&& \vdots &\\
				& \qw & \qw & \qw & \qw & \qw & \qw & \qw & \qw\\
				& \qw & \qw & \qw & \qw & \octrl{1} & \qw & \qw & \qw\\
				\lstick{$\ket{0}_{a_1}$} & \gate{H} & \octrl{-7} & \qw & \octrl{-7} & \targ{} & \octrl{1} & \qw & \qw\\
				\lstick{$\ket{0}_{a_2}$} & \qw & \qw & \qw & \qw & \qw & \gate{R_y(2\arccos \frac{k_1}{k_2})} & \qw & \qw
			\end{quantikz}
		};
	\end{tikzpicture}
\end{equation*}
Again, addition of control qubits to $U_2^\dagger U_1$ does not require the Hadamard gates, the $X$ gates, and the multiple CNOT gates to be controlled.
As a result, the block encoding unitary $U_2^\dagger U_1$ can be qubitized into a $(2n+6)$-qubit circuit $U_H^\ast$
(including two ancillary qubits for the multi-controlled gate and also for the adder) as
\begin{equation*}
	\begin{tikzpicture}
		\node[scale=0.9] at (0,0) {
			\begin{quantikz}[row sep={20pt,between origins}, column sep=10pt]
				\lstick{$\ket{1}_{\mathrm{sgn}}$} & \qw & \qw & \qw & \qw & \qw & \qw & \qw & \qw & \qw & \qw & \qw & \qw & \qw & \gate{Z} & \gate{Z} & \qw\\
				& \qw & \octrl{9} & \octrl{8} & \qw & \qw & \qw & \qw & \qw & \ctrl{8} & \ctrl{9} & \qw & \targ{} & \gate{H} & \octrl{-1} & \gate{H} & \qw\\
				\lstick[4]{$\ket{0}$} & \qw & \qw & \qw & \qw & \targ{} & \gate[wires=8, nwires={2,6}]{\ \ \text{adder}\ \ } & \targ{} & \qw & \qw & \qw & \qw & \qw & \qw & \qw & \qw & \qw\\
				& \vdots &&&& \vdots && \vdots &&&&&&&& \vdots &\\
				& \qw & \qw & \qw & \qw & \targ{} & \qw & \targ{} & \qw & \qw & \qw & \qw & \qw & \qw & \qw & \qw & \qw\\
				& \qw & \qw & \qw & \targ{} & \qw & \qw & \qw & \targ{} & \qw & \qw & \qw & \qw & \qw & \qw & \qw & \qw\\
				\lstick[4]{$\ket{x}_{s}$} & \qw & \qw & \qw & \qw & \qw & \qw & \qw & \qw & \qw & \qw & \qw & \qw & \qw & \qw & \qw & \qw\\
				& \vdots &&&&&&&&&&&&&& \vdots &\\
				& \qw & \qw & \qw & \qw & \qw & \qw & \qw & \qw & \qw & \qw & \qw & \qw& \qw & \qw & \qw & \qw\\
				& \qw & \qw & \octrl{1} & \qw & \qw & \qw & \qw & \qw & \octrl{1} & \qw & \qw & \qw& \qw & \qw & \qw & \qw\\
				\lstick{$\ket{0}_{a_1}$} & \gate{H} & \octrl{1} & \targ{} & \qw & \octrl{-7} & \qw & \octrl{-7} & \qw & \targ{} & \octrl{1} & \gate{H} & \qw& \qw & \octrl{-9} & \qw & \qw\\
				\lstick{$\ket{0}_{a_2}$} & \qw & \gate{R_y} & \qw & \qw & \qw & \qw & \qw & \qw & \qw & \gate{R_y^\dagger} & \qw & \qw& \qw & \octrl{-1} & \qw & \qw
			\end{quantikz}
		};
	\end{tikzpicture}
\end{equation*}
where the latter half of (controlled-)$U_2^\dagger U_1$ and the former half of (controlled-)$U_1^\dagger U_2$ which do not involve control qubits are canceled out,
and the two controlled-adders are combined into a single (non-controlled) adder.
Recalling that a controlled adder uses $4n$ Toffoli gates and
controlling the CNOT gates following Appendix \ref{subsec:Multi-controlled} (instead of turning each into a Toffoli gate),
a controlled-$U_H^\ast$ can be realized using only $(4n+24)$ Toffoli gates.

\begin{table}[h]
	\centering
	\begin{tabular}{c|wc{10mm}wc{50mm}wc{35mm}wc{20mm}}
		\hline
		order of matrix & $d_{\text{code}}$ & \# of logical/physical qubits & \# of Toffoli gates & runtime\\
		\hline
		$2^{32\phantom{0}}$ & $23$ & $198$ / $\sim 3.1\times 10^5$ & $\sim 2.5 \times 10^6$ & $\sim 3$ min.\\
		$2^{64\phantom{0}}$ & $23$ & $336$ / $\sim 4.6\times 10^5$ & $\sim 4.6 \times 10^6$ & $\sim 5$ min.\\
		$2^{128}$ & $25$ & $608$ / $\sim 8.6\times 10^5$ & $\sim 8.8 \times 10^7$ & $\sim 9$ min.\\
		\hline
	\end{tabular}
	\caption{
		Summary of very rough estimates of the costs necessary to solve the problems of Example 2
		using the above fine-tuned circuit with the same settings as in Sec.\,\ref{sec:resource_estimation}.
	}
\end{table}

Fortunately, this seemingly-drastic shortcut still retains flexibility to carry out additional exceptional operations described in Sec.\,\ref{subsec:O_H}.
For example, rotation gates controlled not only on the $a_1$ register but also on the $s$ register enables modifying each non-zero element,
while expanding the $a_1$ register allows (at least conceptually) straightforward generalization of the construction to denser matrices,
where the amount of necessary resources approach to those of the naive implementation as the matrix becomes denser and denser.

\newpage

\renewcommand{\baselinestretch}{1.1}
\bibliographystyle{ytamsalpha}
\bibliography{bib}

\end{document}